\documentclass[aps,prc,floatfix,groupedaddress,showpacs,amsfonts]{revtex4}
\usepackage{bm}
\usepackage{epsfig}
\usepackage{amsmath}
\usepackage[toc,page]{appendix}

\begin{document}

\title{Connection between asymptotic normalization coefficients and resonance widths of mirror states }

\author{A.\,M.~Mukhamedzhanov}
\affiliation{Cyclotron Institute, Texas A\&M University,
College Station, TX 77843}

\begin{abstract}
Asymptotic normalization coefficients (ANCs) are fundamental nuclear constants playing important role in nuclear reactions, nuclear structure and nuclear astrophysics. In this paper a connection between ANCs and resonance widths of the mirror states is established. Using Pinkston-Satchler equation the ratio for  resonance widths and ANCs of mirror nuclei is obtained in terms of the Wronskians from the radial overlap functions and regular solutions of the two-body Schr\"odinger equation with the short-range interaction excluded. This ratio allows one to use microscopic overlap functions for mirror nuclei in the internal region, where they are the most accurate, to correctly predict the ratio of the resonance widths and ANCs for mirror nuclei, which determine the amplitudes of the tails of the overlap functions. If the microscopic overlap functions are not available one can express the Wronskians for the resonances and mirror bound states in terms of the corresponding mirror two-body potential-model wave functions. A further simplification of the Wronskians ratio leads to the equation for the ratio of the resonance widths and mirror ANCs, which is expressed in terms of the ratio of the two-body Coulomb scattering wave functions at  the resonance energy and at the binding energy [N. K. Timofeyuk, R. C. Johnson, and A. M. Mukhamedzhanov, Phys. Rev. Lett. {\bf 91}, 232501  (2003].
In this paper calculations of the ratios of resonance widths and mirror ANCs  for different nuclei are presented. From this ratio one can determine the resonance width if the mirror ANC is known and vice versa. Comparison with available experimental ratios are done.
\end{abstract}

\pacs{21.10.Jx, 21.60.De,25.40.Ny, 24.10.-i}

\maketitle

\section{Introduction}
The asymptotic normalization coefficient (ANC) is a fundamental nuclear characteristics of bound states \cite{blokh77,blokhintsev84} playing an important role in nuclear reaction and structure physics. The ANCs determine the normalization of the peripheral part of transfer reaction amplitudes \cite{blokh77,blokhintsev84} and overall normalization of the peripheral radiative capture processes \cite{muktim90,muk90,muk2001,muk2011}.  
In the R-matrix approach the ANC determines the normalization of the external nonresonant radiative capture amplitude and the channel radiative reduced width amplitude \cite{muktr99,tang,kroha2011}.
In \cite{timofeyuk2003,muk2012} relationships between mirror proton and neutron ANCs were obtained.

However, the ANCs are important characteristics not only of the bound states but also resonances, see
\cite{muktr99}. The width of a narrow resonance can be expressed in terms of the ANC of the Gamow wave function or of the $R$-matrix resonant outgoing wave.
Because the ANC of a narrow resonance state is related to the resonance width of this state the relationship between the ANCs of mirror states \cite{timofeyuk2003,muk2012} can be extended to the relationship between the ANCs of the bound states and resonance widths of the mirror states. The first such attempt was done in \cite{timofeyuk2003}. In this paper  the relationship between the ANCs and resonance widths is established  based on the Pinkston-Satchler equation used in \cite{muk2012} for the ANCs of the mirror bound states. 
The obtained ratio of the resonance width and the ANC of the mirror bound state is expressed  in terms of the ratio of the Wronskians containing the overlap functions of the mirror 
resonance and bound states in the internal region where these overlap functions can be calculated quite accurately using {\it ab initio} approach. If these overlap functions are not available, as an approximation they can be replaced by the mirror resonance and bound state wave functions calculated using the two-body potential model. 
Assuming that the mirror resonant and bound-state wave functions are identical  in the nuclear interior one can replace the Wronskian ratio for the  resonance width and  the
ANC of the mirror bound state  by the equation derived  in \cite{timofeyuk2003}, which does not require a knowledge of the internal resonant and bound-state wave functions.  

Connection between the ANC and the resonance width of the mirror resonance state provides a powerful indirect method to obtain information,
which is unavailable directly. If, for instance, the resonance width is unknown it can be determined through the known ANC of the mirror state and vice versa. For example, near the edge of the stability valley neutron binding energies become so small, that the mirror proton states are resonances.  Using the relationship between the mirror resonance width 
and the ANC the resonance width can be determined. Also  loosely bound states $\alpha+ A$ become resonances in the mirror
nucleus $\alpha+ B$, where charge $Z_{B}e > Z_{A}e$. Using the method  developed here  one can find one of the missing quantities,  the resonance width of the narrow resonance state or the mirror ANC. 
In what follows the system of units in which $\hbar=c=1$ is used throughout the paper. 
\\

\subsection{ANC in the scattering theory and Schr\"odinger formalism}
\label{ANCscth1}

The ANC enters the theory in two ways \cite{blokh77}. In the scattering theory the residue at the poles of the elastic scattering $S$ matrix corresponding to bound states 
can be expressed in terms of the ANC:
\begin{align}
S_{l_{B}\,j_{B};l_{B}\,j_{B}}^{J_{B}} \xrightarrow{k_{aA} \to k_{aA}^{bs}}\,\frac{A_{l_{B}\,j_{B}}}{k - i\,\kappa_{aA}}    
\label{elscatSm1}
\end{align}
with the residue
\begin{align}
A_{l_{B}\,j_{B}}^{J_{B}}= -i^{2\,l_{B} + 1}\,e^{i\,\pi\,\eta_{aA}^{bs}}\,\big(C^{B}_{aA\,l_{B}\,j_{B}\,J_{B}}\big)^{2}.
\label{residue1}
\end{align}
Here, $\,C^{B}_{aA\,l_{B}\,j_{B}\,J_{B}}$ is the ANC for the virtual decay  of the bound state $B(aA)$  in the channel with the relative orbital angular momentum $\,\,l_{B}$ of $a$ and $A$, the total angular momentum $\,\,j_{B}$ of $a$ and total angular momentum $\,\,J_{B}$ of the system $a + A$,  $k_{aA}$ is the relative momentum of particles $a$ and $A$.
\begin{align}
\eta_{aA}^{bs}=\frac{Z_{a}\,Z_{A}\,e^{2}\,\mu_{aA}}{\kappa_{aA}}
\label{eta01}
\end{align}
is the Coulomb parameter for the bound state $B=(a\,A)$,  $\;\kappa_{aA}= \sqrt{2\,\mu_{aA}\,\varepsilon_{aA}}$ is the bound-state wave number, $\,\varepsilon_{aA}= m_{a} + m_{A} - m_{B}$ is the binding energy for the virtual decay $B \to a + A$, $Z_{i}\,e$ and $m_{i}$ is the charge and mass of particle $i$, and $\mu_{aA}$ is the reduced mass of $a$ and $A$.  Note that the singling out the factor
$e^{i\,\pi\,\eta_{aA}^{bs}}$ in the residue makes the ANC for bound states real.
 
Equations (\ref{elscatSm1})  and  (\ref{residue1}), which  were proved for the bound states in  \cite{kramers,heisenberg, meller, hu, zeldovich, perelomov},
 can be extended for resonance states.   
  
 \section{Connection between ANC and  resonance width}
\label{ConnectionANCReswidth}

The proof  of   the connection between the residue in the resonance pole of the elastic scattering $S$ matrix and the ANC of the resonance state is not trivial.  In this section  is presented a general proof of the connection of the residue in the pole of the $S_{l_{B}}(k_{aA})$ matrix element  with the ANC , which is valid both for the bound states and resonances. The  potential  is given by the sum of the short-range nuclear plus the long-range Coulomb potentials. 
Taking into account that the residue of the elastic scattering $S$ matrix  in the resonance pole is expressed  in terms of the resonance width, one can obtain a connection between the ANC and the resonance width.

 Let me consider two spinless particles $a$ and $A$ with relative momentum $k_{aA}^{2}= 2\,\mu_{aA}\,E_{aA}$, relative energy $E_{aA}$ and the reduced mass $\mu_{aA}$ in the partial wave $l_{B}$ at which the system $B=a+A$ has a resonance or a bound state.
The radial wave function $\psi_{k_{aA}l_{B}}(r) =\frac{u_{k_{aA}l_{B}}(r)}{r}$ satisfies the Schr\"odinger equation  in the partial wave $l_{B}$:
\begin{align}
\frac{{\partial}^{2} u_{k_{aA}l_{B}}(r) }{{\partial {r}^{2}}}+ \big[k_{aA}^2 - 2{\mu _{aA}}V(r)\,\, - \,\frac{{l_{B}(l_{B} + 1)}}{{{r^2}}}\big]u_{k_{aA}l_{B}}(r) = 0.
\label{Schreq1}
\end{align}
Here $V(r)= V^{N}(r) + V^{C}(r)$, $\,V^{N}(r)$ is the short-range nuclear potential and $V^{C}(r)$ is the long-range Coulomb one.
For potentials satisfying the condition $\mathop {\lim }\limits_{r \to 0 }r^{2}\,V(r) \to 0$
\begin{align}
u_{k_{aA}l_{B}}(r) \sim r^{l_{B}+1}, \qquad  r \to 0.
\label{ur01}
\end{align}

Now one should take the derivative over $k_{aA}$ from the left-hand-side of Eq. (\ref{Schreq1}), multiply the result by $u_{k_{aA}l_{B}}(r)$ and subtract from it Eq. (\ref{Schreq1}) multiplied by $\partial {u_{k_{aA}l_{B}}}(r)/\partial k_{aA}$.
Integrating the obtained expression from $r=0$ until $r=R$ and taking into account Eq. (\ref{ur01})
one gets
\begin{align}
\int\limits_0^R {dr\,u_{k_{aA}l_{B}}^2(r) = \,\frac{1}{{2k_{aA}}}\left[ {\frac{{\partial {u_{k_{aA}l_{B}}}(R)}}{{\partial k_{aA}}}\frac{{\partial {u_{k_{aA}l_{B}}}(R)}}{{\partial R}} - {u_{k_{aA}l_{B}}}(R)\frac{{{\partial ^2}{u_{k_{aA}l_{B}}}(R)}}{{\partial k_{aA}\,\partial R}}} \right]}. 
\label{derkR1}
\end{align}

Taking $R$ so large that $u_{k_{aA}l_{B}}(R)$ can be replaced by its leading asymptotic term one gets
\begin{align}
{u_{k_{aA}l_{B}}}(R) \approx {{\tilde C}_{l_{B}}}\left[ {{e^{i\rho }} - {{( - 1)}^l_{B}}S_{l_{B}}^{ - 1}(k_{aA}){e^{ - i\rho }}} \right],
\label{uklas1}
\end{align}
where $\rho  = k_{aA}R - \eta_{aA}\, \ln \,2k_{aA}R,\,\,\,\,\,\,\,\eta_{aA}  = ({Z_a}{Z_A}/137){\mu _{aA}}/k_{aA}\,$ is the Coulomb parameter of the $a+A$ system.
\begin{align}
S_{l_{B}}(k_{aA}) = e^{2\,i\,[\delta_{l_{B}}^{C}(k_{aA}) + \delta_{l_{B}}^{CN}(k_{aA})]}
\label{Sl1}
\end{align}
is the elastic scattering $S$-matrix element,  
$\delta_{l_{B}}^{C}(k_{aA})$ and $\delta^{CN}_{l_{B}}(k_{aA})$ are the Coulomb and Coulomb-modified nuclear scattering phase shifts in the $l_{B}$-th partial wave.  ${\tilde C}_{l_{B}}$ does not depend on $k_{aA}$ and can be determined from the normalization condition of the bound or resonance state wave functions.

Assume now that the elastic scattering $S_{l_{B}}(k_{aA})$-matrix element  has a first order pole at $k_{aA}
=k_{p}$ with the residue $A_{l_{B}}$ corresponding to the bound state  $k_{p}= i\,\kappa_{aA}$ and to the resonance state $k_{p}=k_{aA(R)}=k_{aA\,(0)} - {\rm Im}k_{aA\,(R)},$ where $\;k_{aA\,(0)}= {\rm Re}k_{aA\,(R)}$:
\begin{align}
S_{l_{B}}(k_{aA})=   \frac{A_{l_{B}}}{k_{aA}-k_{p} } + g_{l_{B}}(k_{AA}),
\label{Spole1}
\end{align}
$g_{l_{B}}(k_{aA})$ is a regular function at $k_{aA}=k_{p}$.

Substituting Eqs (\ref{uklas1}) and (\ref{Spole1})  into the right-hand-side  of Eq. (\ref{derkR1})  and performing the differentiation over $k_{aA}$ and $R$ and taking $k_{aA}=k_{p}$ one gets
\begin{align}
\int\limits_{0}^{R} {\rm d}r\,u_{k_{p}l_{B}}^2(r) = i\,(-1)^{l_{B}+1}{\tilde C}_{l_{B}}^{2}/A_{l_{B}}  - \frac{i}{2\,k_{p}}\,e^{2\,i\,\rho_{0}}.
\label{primeeq1}
\end{align} 

On the left-hand-side under the integral sign we have the function $u_{k_{p}l_{B}}^2(r)$, which is regular at $r=0$ (see Eq. (\ref{ur01}) and  satisfies the radiation condition:
\begin{align}
u_{k_{p}l_{B}}(r)  \stackrel{r >R_{N}}{=}  C_{l_{B}}\,W_{-i\,\eta_{p},\,l_{B}+ 1/2}(-2\,i\,k_{p}\,r).
\label{radcond1}
\end{align}
Here $R_{N}$ is the $a-A$ nuclear interaction radius. 
For the  bound state $k_{p}= i\,\kappa_{aA}$   and
\begin{align}
u_{i\,\kappa_{aA}l_{B}}(r)  \stackrel{r >R_{N}}{=}  C_{l_{B}}\,W_{-\eta_{aA}^{bs},\,l_{B}+ 1/2}(2\,k_{p}\,r)     \stackrel{r \to \infty}{\approx}    C_{l_{B}} e^{-\kappa_{aA}\,r - \eta_{aA}^{bs}\,\ln(2\,\kappa_{aA}\,r)}.
\label{radcondbs1}
\end{align}
For the resonance state  $k_{p}= k_{aA(R)} $   and  $u_{k_{aA(R)}l_{B}}(r)$ is the resonance  Gamow wave function with the resonance energy $E_{aA(R)}$:
\begin{align}
&u_{k_{aA(R)}l_{B}}(r)  \stackrel{r >R_{N}}{=}  C_{l_{B}}\,W_{-i\,\eta_{aA}^{R},\,l_{B}+ 1/2}(-2\,i\,k_{aA(R)}\,r)     \stackrel{r \to \infty}{\approx}    e^{-\pi\,\eta_{aA}^{R}/2}\, C_{l_{B}} e^{ i\,k_{aA(R)}\,r -i\, \eta_{aA}^{R}\,\ln(2\,k_{aA(R)}\,r)}                                                                                     \nonumber\\
&= {\tilde C}_{l_{B}} e^{ i\,k_{aA(R)}\,r -i\, \eta_{aA}^{R}\,\ln(2\,k_{aA(R)}\,r)} .
\label{radcondres1}
\end{align}
Here, $ \eta_{aA}^{R}=\,\frac{Z_{a}\,Z_{A}\,e^{2}\,\mu_{aA}}{k_{aA\,(R)}}$ the  $a-A$ Coulomb parmeter of the resonance.  For the resonance  
\begin{align}
{\tilde C}_{l_{B}}= e^{-\pi\,\eta_{aA}^{R}/2}\,C_{l_{B}}.  
\label{res1}
\end{align}
Thus  one can use two equivalent definitions of the ANC, which differ by a factor of  $e^{-\pi\,\eta_{aA}^{R}/2}$ for the resonance state.
Formally, one can use two definitions of the ANC for the bound states:
\begin{align}
 {\tilde C}_{l_{B}}= e^{i\,\pi\,\eta_{aA}^{bs}/2}\,C_{l_{B}}.   
\label{bs1}
\end{align}
However, the $C_{l_{B}}$, which is real for the bound states, is the standard definition of the ANC for the bound states and will be used in this paper for the bound states.

Also  the following definitions are used:
\begin{align}
&E_{aA(R)}=  k_{aA(R)}^{2} /(2\,\mu_{aA}) = E_{aA(0)} - i\,\Gamma_{aA}/2, \qquad         E_{aA(0)}= \big[k_{aA(0)}^{2}- ({\rm Im}k_{aA(R)})^{2}\big]/(2\,\mu_{aA}),  \nonumber\\
&\Gamma_{aA}= 2\,k_{aA(0)}\,{\rm Im}k_{aA(R)}/\mu_{aA}.
\label{ERkR1}
\end{align} 

 For the bound states  the asymptotic of the bound state wave is exponentially  decaying and the bound-state wave function can be normalized.
For the resonance state the Gamow wave function asymptotically oscillates and exponentially  increasing.  To normalize the Gamow wave function  one can use Zeldovich 
regularization procedure \cite{zeldovich} which is a particular case of the more general Abel regularization:
\begin{align}
\lim\limits_{\beta \to +0}\,\int\limits_{0}^{\infty} \,{\rm d}r e^{-\beta\,r^{2}}\,u_{k_{aA(R)}l_{B}}^{2}(r) =1.
\label{normGamow1}
\end{align}

For the bound state one can take under the integral sign $\beta=0$ and  obtain the usual normalization procedure.  For the resonance state one can take the limit $\beta \to 0$ only
after performing the integration over $t$. Note that Zel'dovich normalization was introduced for exponentially decaying potentials. 
In Appendix A is shown that Zel'dovich regularization procedure  works even for the Coulomb potentials. 

For any finite $R\,$  one  can rewrite  Eq. (\ref{normGamow1})  as
\begin{align}
\int\limits_{0}^{R} \,{\rm d}r\,u_{k_{aA(R)}}^{2}(r)  + \lim\limits_{\beta \to +0}\,\int\limits_{R} ^{\infty}\,{\rm d}r e^{-\beta\,r^{2}} \,u_{k_{aA(R)}l_{B}}^{2}(r) =1.
\label{normsplit1}
\end{align}

Assume that $R$ is so large that one can use the asymptotic expression  (\ref{radcond1})  and Eq. (A.5) of Appendix.  It leads to 
\begin{align}
\int\limits_{0}^{R} \,{\rm d}r\,u_{k_{aA(R)}l_{B}}^{2}(r) =1 - \frac{i}{2\,k_{aA(R)}}{\tilde C}_{l_{B}}^{2}\,e^{2\,i\,\rho_{0}}.
\label{normGamow2}
\end{align}

Comparing Eqs (\ref{primeeq1}) and (\ref{normGamow2})  one arrives to the final equation, which expresses the residue in the pole of the elastic scattering $S$-matrix  in terms of the ANC:
\begin{align}
A_{l_{B}}=  -i^{2\,l_{B} + 1}\,{\tilde C}_{l_{B}}^{2}.
\label{AlCl1}
\end{align}
Equation (\ref{AlCl1}) is universal  and  valid for  bound state poles and resonances.
In terms of the standard  ANC $C_{l_{B}}$ the residue in the resonance pole is
\begin{align}
A_{l_{B}}=  -i^{2\,l_{B} + 1}\,e^{-\pi\,\eta_{aA}^{R}}\,C_{l_{B}}^{2}
\label{AltildeCstandCl1}
\end{align}
and for the bound state is given by Eq. (\ref{residue1}).

 Now it will be shown how to relate the ANC ${\tilde C}_{l_{B}}$ to the resonance width $\Gamma_{aA}$ . 
To this end one can write 
\begin{align}
S_{l_{B}}(k_{aA})= e^{2\,i\,\delta_{l_{B}}^{pot}}\,\frac { (k_{aA} + k_{p})( k_{aA} - k_{p}^{*} ) }{(k_{aA} - k_{p})( k_{aA} +    k_{p}^{*}  ) }, 
\label{Slk1}
\end{align}
where $\,\delta_{l_{B}}^{pot}$  is the non-resonant scattering phase shift.  At $k_{p}= k_{aA(R)}$  and at  $k_{aA} \to k_{aA(R)}$  
\begin{align}
A_{l_{B}}(k_{aA}) = -2\,i\,k_{aA(R)}\,\gamma\,\big[(1+ \gamma^{2})^{1/4} +\,(1+ \gamma^{2})^{-1/4} ]^{-1}\,e^{i[2\,\delta_{l_{B}}^{pot}(k_{aA(R)})  - 1/2 \,arctan (\gamma) ]},
\label{Al2}
\end{align}
$\gamma= \frac{\Gamma_{aA}}{2\,E_{aA(0)}}$.
Equation  (\ref{Al2})  expresses the residue of the $S$-matrix elastic scattering element  in terms of the resonance energy and the resonance width for broad resonances.

Recovering now all the quantum numbers one gets for a narrow resonance  ($\gamma <<1$)  up to terms of order $\sim \gamma$ 
\begin{align}
\big({\tilde C}^{B}_{aA\,l_{B}\,j_{B}\,J_{B}}\big)^{2}= (-1)^{l_{B}}\,e^{i\,2\,\delta_{l_{B}\,j_{B}\,J_{B}}^{p}(k_{aA(0)})}\,\frac{\mu_{aA}\,\Gamma_{aA\,l_{B}\,j_{B}\,J_{B}}}{k_{aA(0)}},
\label{ANCGamma1}
\end{align} 
where  $\Gamma_{aA\,l_{B}\,j_{B}\,J_{B}}$ is the resonance width, $\,\delta_{l_{B}\,j_{B}\,J_{B}}^{p}(k_{aA}^{0})$ is the potential (non-resonance) scattering phase shift at the real resonance relative momentum $\,k_{aA\,(0)}$.
 This equation is my desired equation, which relates the ANC  of the narrow resonance to the resonance width.
 
 The  residue in the resonance pole with recovered all the quantum numbers  is
 \begin{align}
A_{l_{B}\,j_{B}}^{J_{B}}= -i^{2\,l_{B} + 1}\,({\tilde C}_{aA\,l_{B}\,j_{B}\,J_{B}}^{B})^{2}.
\label{residueR1}
\end{align}

For the Breit-Wigner resonance  (${\rm Im}k_{aA\,(R)} << {\rm Re}k_{aA\,(R)}= k_{aA\,(0)}$)  Eq. (\ref{residueR1}) takes the form 
\begin{align}
A_{l_{B}\,j_{B}}^{J_{B}}= -i^{2\,l_{B} + 1}\,e^{-\,\pi\,\eta_{aA}^{(0)}}\,\big(C^{B}_{aA\,l_{B}\,j_{B}\,J_{B}}\big)^{2} =-i^{2\,l_{B} + 1}\,({\tilde C}_{aA\,l_{B}\,j_{B}\,J_{B}}^{B})^{2},
\label{residueresBWR1}
\end{align}
where $\eta_{aA}^{(0)}= Z_{a}\,Z_{A}\,e^{2}\,\mu_{aA}/k_{aA\,(0)}$. 
In terms of the resonance width  the residue of the elastic scattering $S$-matrix element in the resonance pole is 
\begin{align}
A_{l_{B}j_{B}}^{J_{B}}= -\,ie^{2\,i\,\delta_{l_{B}\,j_{B}\,J_{B}}^{p}(k_{aA}^{0})}\,\frac{\mu_{aA}}{k_{aA\,(0)}}\,\Gamma_{aA\,l_{B}j_{B}J_{B}}.
\label{residreswidth1}
\end{align}

 \section{ANCs and the overlap functions}
 
Equations obtained in the previous section, which express the residues of the $S$-matrix elastic element  in terms of the ANCs of the bound states and resonances, 
provide the most general and model-independent definition of the ANCs.  From other side, in the Schr\"odinger formalism of the wave functions the ANC is defined as the amplitude of the tail of the overlap function of the bound state wave functions of $B,\,A$ and $a$. The overlap function is given by 
\begin{align}
&I_{aA}({\rm {\bf r}}_{aA}) = \, <\psi_{c}\,|\,{\varphi _B}({\xi _A},\,{\xi _a},\,{\rm {\bf r}}_{aA})>       \nonumber\\
&= \,\sum\limits_{{l_B}{m_{{l_B}}}{j_B}{m_{{j_B}}}} { < {J_A}{M_A}\,\,{j_B}{m_{{j_B}}}|{J_B}{M_B} >  < {J_a}{M_a}\,{l_B}{m_{{l_B}}}|{j_B}{m_{{j_B}}} > } \,{Y_{{l_B}{m_{{l_B}}}}}({\rm {\bf {\widehat r}}}_{aA})\,I_{aA\,\,{l_B}{j_B}\,J_{B}}({r_{aA}}). 
\label{overlapfunction1}
\end{align} 
Here 
\begin{align}
\psi_{c}= \sum\limits_{m_{{j_B}}{m_{{l_B}}}{M_A}{M_{a}}} { < {J_A}{M_A}\,\,{j_B}{m_{{j_B}}}|{J_B}{M_B} >  < {J_a}{M_a}\,{l_B}{m_{{l_B}}}|{j_B}{m_{{j_B}}} > }\,{\widehat A_{aA}}\{ \varphi_{A}(\xi_{A})\,\varphi_{a}({\xi _a})\,Y_{ {l_B}\,{m_{l_{B}}} }({\rm {\bf {\widehat r}}}_{aA})\} 
\label{channelwf1}
\end{align}
is the two-body $a+ A$ channel wave function in the $jj$ coupling scheme, $<j_{1}\,m_{1}\,\,j_{2}\,m_{2}|j\,m>$ is the Clebsch-Gordan coefficient, ${\widehat A_{aA}}$ is the antisymmetrization operator between the nucleons of nuclei $a$  and $A$; $\,\,\varphi_{i}(\xi_{i})$ represents the fully antisymmetrized bound state wave function of nucleus $i$ with $\xi_{i}$ being a set of the internal coordinates including spin-isospin variables, $\,\,J_{i}$ and $M_{i}$ are the spin and its projection of nucleus $i$. Also ${\rm {\bf r}}_{aA}$ is the radius vector connecting the centers of mass of nuclei $a$ and $A$, $\,\,{\rm {\bf {\hat r}}}_{aA} = \,{\rm {\bf r}}_{aA}/r_{aA}$, $\,\,Y_{l_B\,m_{l_{B}}}({\rm {\bf {\hat r}}}_{Aa})$ is the spherical harmonics, and  $I_{aA\,\,{l_B}{j_B}J_{B}}({r_{Aa}})$ is the radial overlap function. The summation over ${l_B}$ and ${j_B}$ is carried out over the values allowed by the angular momentum and parity conservation in the virtual process $B \to A + a$.

The radial overlap function is given by
\begin{align}
&I_{aA\,\,{l_B}{j_B}\,J_{B}}({r_{aA}})\, = \, <{\hat A}_{aA}\,\{{\varphi _A}({\xi _A})\,{\varphi _a}({\xi_a})\,Y_{ {l_B}\,{m_{l_{B}}} }({\rm {\bf {\widehat r}}}_{aA})\}|\,{\varphi _B}({\xi _A},\,{\xi _a};\,{\rm {\bf r}}_{aA})>                                                 \nonumber\\
&= \left( \begin{gathered}
  A \hfill \\
  a \hfill \\ 
\end{gathered}  \right)^{\frac{1}{2}}<{\varphi _A}({\xi _A})\,{\varphi _a}({\xi_a})\,Y_{ {l_B}\,{m_{l_{B}}} }({\rm {\bf {\widehat r}}}_{aA})|\,{\varphi _B}({\xi _A},\,{\xi _a};\,{\rm {\bf r}}_{aA})>.
\label{radoverlap1}
\end{align}
Eq. (\ref{radoverlap1}) follows from a trivial observation that, because ${\varphi _B}$ is fully antisymmetrized, the antisymmetrization operator ${\hat A_{aA}}$ can be replaced by the factor  
$\left( \begin{gathered}
  A \hfill \\
  a \hfill \\ 
\end{gathered}  \right)^{\frac{1}{2}}$.
In what follows, in contrast to Blokhintsev et al (1977), we absorb this factor into the radial overlap function.     

The tail of the radial overlap function (${r_{aA}} > \,R_{aA}$) in the case of the normal asymptotic behavior is given by
\begin{align}
I_{aA\,\,{l_B}\,{j_B}\,J_{B}}({r_{aA}})\,\,=\,\,{C^{B}_{aA\,{l_B}\,{j_B}\,J_{B}}}\frac{{{W_{ - {\eta _{aA}^{bs}},\,\,{l_B} + 1/2}}(2\,{\kappa_{aA}}{r_{aA}})}}{{{r_{aA}}}}\xrightarrow{{{r_{aA}} \to \infty }}{C^{B}_{aA\,{l_B}{j_B} J_{B}}}\,\frac{{{e^{ - \kappa _{aA}{r_{aA}} - {\kern 1pt} \eta _{aA}^{bs}\ln (2\,\kappa _{aA}{r_{aA}})}}}}{{{r_{aA}}}}.
\label{asymptoverlap1}
\end{align}

Formally the radial resonance overlap function for the Breit-Wigner resonance in the external region ($r_{aA} > R_{aA}$)  can be obtained  from  Eq.  (\ref{asymptoverlap1}) by the substitution $\kappa_{{aA}  }=-i\,k_{aA\,(0) }$:
\begin{align}
&I_{aA\,\,{l_B}\,{j_B}\,J_{B}}(k_{aA(0)},\,{r_{aA}}) =\,\,{C^{B}_{aA\,{l_B}\,{j_B}\,J_{B}}}\frac{{{W_{ - i\,{\eta _{aA}^{(0)}},\,\,{l_B} + 1/2}}(-2\,i\,k_{aA\,(0)}{r_{aA}})}}{{{r_{aA}}}} 
\label{OvW1}         \\
&\xrightarrow{{{r_{aA}} \to \infty }}{C^{B}_{aA\,{l_B}{j_B} J_{B}}}\,\frac{{{e^{  i\,k_{aA\,(0)}\,{r_{aA}} - \,i\,\eta _{aA}^{(0)}\ln (-2\,i\,k_{aA\,(0)}\,{r_{aA}})}}}}{{{r_{aA}}}} 
\nonumber\\
&= {\tilde C}^{B}_{aA\,{l_B}{j_B} J_{B}}\,\frac{{{e^{  i\,k_{aA\,(0)}\,{r_{aA}} - \,i\,\eta _{aA}^{(0)}\ln (2\,k_{aA\,(0)}\,{r_{aA}})}}}}{{{r_{aA}}}}.
\label{asymptoverlapres1}
\end{align}

In the $R$-matrix approach the resonant wave function is considered at the real part of the resonance energy $E_{aA\,(0)}$.
The overlap function of the Breit-Wigner resonance state is given by
\begin{align}
&I_{aA}({\rm {\bf r}}_{aA}) = \, <\psi_{c}\,|\,{\Psi}({\xi _A},\,{\xi _a};\,{\rm {\bf r}}_{aA})>       \nonumber\\
&= \,\sum\limits_{{l_B}{m_{{l_B}}}{j_B}{m_{{j_B}}}} { < {J_A}{M_A}\,\,{j_B}{m_{{j_B}}}|{J_B}{M_B} >  < {J_a}{M_a}\,{l_B}{m_{{l_B}}}|{j_B}{m_{{j_B}}} > } \,{Y_{{l_B}{m_{{l_B}}}}}({\rm {\bf {\widehat r}}}_{aA})\,I_{aA\,\,{l_B}{j_B}\,J_{B}}(k_{aA(0)},\,{r_{aA}}). 
\label{overlapfunction1}
\end{align} 
Here, $I_{aA\,\,{l_B}{j_B}\,J_{B}}(k_{aA(0)},\,r_{aA})$ at $r_{aA} > R_{aA}$ for the Breit-Wigner resonance is determined by 
\begin{align}
I_{aA\,\,{l_B}\,{j_B}\,J_{B}}(k_{aA(0)},\,{r_{aA}}) = {{\tilde C}^{B}_{aA\,{l_B}\,{j_B}\,J_{B}}}\,i^{l_{B}}\,\frac{ O_{l_{B}}(k_{aA(0)},r_{aA})}{r_{aA}} = \sqrt{\frac{\mu_{aA}}{k_{aA(0)}}\,\Gamma_{aA\,l_{B}j_{B}J_{B}}}\,e^{-i\,\delta_{l_{B}}^{hs}}\,\frac{O_{l_{B}}(k_{aA(0)},r_{aA})}{r_{aA}},
\label{Rmatrixasympt1}
\end{align}
\begin{align}
O_{l_{B}}(k_{aA},r_{aA})= iF_{l_{B}}(k_{aA},r_{aA}) + G_{l_{B}}(k_{aA},r_{aA}) = e^{i\,\delta_{l_{B}}^{hs}}\,\sqrt{
F_{l_{B}}^{2}(k_{aA},r_{aA}) + G_{l_{B}}^{2}(k_{aA},r_{aA})}\,,
\label{OlFG1}
\end{align}
$F_{l_{B}}(k_{aA},r_{aA})$ and $G_{l_{B}}(k_{aA},r_{aA})$ are the regular and singular Coulomb solutions.
Note that in the $R$-matrix method the potential scattering phase shift is given by the hard-sphere scattering phase shift $-\delta_{l_{B}}^{hs}$. 

\section{Connection between  Breit-Wigner resonance width  and  ANC of  mirror resonance and bound states  from Pinkston-Satchler equation}

In \cite{muk2012} the relationship between the mirror proton and neutron ANCs was derived using the Pinkston-Satchler equation \cite{pinkston,philpott}. Here I extend this derivation to obtain the ratio for the resonance width and the ANC of the mirror bound state in terms of the Wronskians, which follows from the Pinkston-Satchler equation. 

First, using the Pinkston-Satchler equation I derive the equation for the ANC of the narrow resonance state, which contains the source term \cite{muk90,tim1998}. This derivation is valid for both bound and resonance state. That is why following \cite{muk2012} 
I start from the Schr\"odinger equation for the resonance scattering wave function at the real part E $_{aA(0)}$ of the resonance  energy :
\begin{align}
(E_{(0)} - {\widehat T_A} - {\widehat T_a} - {\widehat T_{aA}} - \,V_{a}  - \,V_{A} - V_{aA} ){\Psi}({\xi _A},{\xi _a};\,{\rm {\bf r}}_{aA}) = 0.                             
\label{Schrequat1}                              
\end{align}
Here, ${\widehat T_i}$ is the internal motion kinetic energy operator of nucleus 
$i$, $\,{\widehat T_{aA}}$ is the kinetic energy operator of the relative motion of nuclei 
$a$ and $A$, $V_{i}$ is the internal potential of nucleus $i$ and $V_{aA}$ is the interaction potential between $a$ and $A$, $\,\,E_{(0)} = E_{aA(0)} - \varepsilon_{a} - \varepsilon_{A}$ is the total energy of the system $a+A$ in the continuum.

Multiplying the Schr\"odinger equation (\ref{Schrequat1})  from the left by  
\begin{align}
\,{\left( \begin{gathered}
  A \hfill \\
  a \hfill \\ 
\end{gathered}  \right)^{1/2}}\,\sum\limits_{{m_{{j_B}}}{m_{{l_B}}}M_{A}M_{a}} { < {J_A}{M_A}\,\,{j_B}{m_{{j_B}}}|{J_B}{M_B} >  < {J_a}{M_a}\,{l_B}{m_{{l_B}}}|{j_B}{m_{{j_B}}} > } \,{Y^{*}_{{l_B}{m_{{l_B}}}}}({\rm {\bf {\widehat r}}}_{aA})\,\varphi_{A}(\xi_{A})\,\varphi_{a}(\xi_{a})
\label{projectaAstate1}
\end{align}
I get the equation for the radial overlap function with the source term \cite{tim1998}
\begin{align}
\Big(E_{aA(0)} - {\hat T}_{r_{aA}}- V^{centr}_{l_{B}} - U_{aA}^{C} \Big)\,I_{aA\,\,{l_B}\,{j_B}\,J_{B}}^{B}(r_{aA}) = Q_{l_{B}j_{B}J_{a}J_{A}J_{B}}(r_{aA}),
\label{radeqst1}
\end{align}
where at $r_{aA} > R_{aA}$ the radial overlap function $I_{aA\,\,{l_B}\,{j_B}\,J_{B}}^{B}(r_{aA})$ is given by Eq. (\ref{Rmatrixasympt1}).
Also ${\hat T}_{r_{aA}}$ is the radial relative kinetic energy operator of the particles $a$ and $A$, $\,V_{l_{B}}^{centr}$ is the centrifugal barrier for the relative motion of $a$ and $A$ with the orbital momentum $l_{B}$;  $\,Q_{l_{B}j_{B}J_{a}J_{A}J_{B}}(r_{aA})$ is the source term
\begin{align}
&Q_{l_{B}j_{B}J_{a}J_{A}J_{B}}(r_{aA}) = \sum\limits_{m_{j_B}m_{l_B}M_{A}M_{a}} { < {J_A}{M_A}\,\,{j_B}{m_{{j_B}}}|{J_B}{M_B} >  < {J_a}{M_a}\,{l_B}{m_{{l_B}}}|{j_B}{m_{{j_B}}} > }\nonumber\\
& \times\,{\left( \begin{gathered}A \hfill \\
a \hfill \\ 
\end{gathered}  \right)^{1/2}}\,\int\,{\rm d}\,\Omega_{{\rm {\bf r}}_{aA}}\,<\varphi_{a}(\xi_{a})\,\varphi_{A}(\xi_{A})|V_{aA} - U_{aA}^{C}|Y_{{l_B}{m_{{l_B}}}}^{*}({\rm {\bf {\widehat r}}}_{aA})\Psi(\xi_{a},\xi_{A};{\rm {\bf r}}_{aA})>.
\label{sourceeqn1}
\end{align}
The integration in the matrix element $<\varphi_{a}(\xi_{a})\,\varphi_{A}(\xi_{A})|V_{aA} - U_{aA}^{C}|Y_{{l_B}{m_{{l_B}}}}^{*}({\rm {\bf {\widehat r}}}_{aA})\Psi(\xi_{a},\xi_{A};{\rm {\bf r}}_{aA})>\,\,$ in Eq. (\ref{sourceeqn1}) is carried out over all the internal coordinates of nuclei $a$ and $A$.  Note that the antisymmetrization operator ${\hat A}_{aA}$ in Eq. (\ref{projectaAstate1}) is replaced by  ${\left( \begin{gathered}A \hfill \\
a \hfill \\ 
\end{gathered}  \right)^{1/2}}$ because the operator $E_{(0)} - {\widehat T_A} - {\widehat T_a} - {\widehat T_{aA}} - \,V_{a}  - \,V_{A} - V_{aA}$ in Eq. (\ref{Schrequat1}) is symmetric over interchange of nucleons of $a$ and $A$, while $\Psi(\xi_{a},\xi_{A};{\rm {\bf r}}_{aA})$ is antisymmetric. 
For charged particles it is convenient to single out the channel Coulomb interaction $U_{aA}^{C}(r_{aA})$ between the centers of mass of nuclei $\,a$ and $A$. 

Owing to the presence of the short-range potential operator $V_{aA} - U_{aA}^{C}$ (potential $V_{aA}$ is the sum of the nuclear $V^{N}_{aA}$ and the Coulomb $V_{aA}^{C}$ potentials and subtraction of $U_{aA}^{C}$removes the long-range Coulomb term from $V_{aA}$) the source term is also a short-range function. 
Then  Eq. (\ref{radeqst1}) can be rewritten as  
\begin{align}
I_{aA\,\,{l_B}\,{j_B}\,J_{B}}(k_{aA(0)},\,r_{aA}) = \,\frac{1}{R_{aA}}\,\int\limits_{0}^{R_{aA}}\,{\rm d}r'_{aA}\,r'_{aA}\,G_{l_{B}}^{C}(r_{aA},\,r'_{aA}; E_{aA(0)})\,Q_{l_{B}j_{B}J_{a}J_{A}J_{B}}(r'_{aA}).
\label{radeqst2}
\end{align}
The partial Coulomb two-body Green function is given by  \cite{newton}
\begin{align}
G_{l_{B}}^{C}(r_{aA},r'_{aA}; E_{aA})= -2\,\mu_{aA}\,\frac
{\varphi_{l_{B}}^{C}(k_{aA},\,r_{aA\,<})\,f_{l_{B}}^{C(+)}(k_{aA},\,r_{aA\,>})}{L_{l_{B}}^{C(+)}},
\label{Gredenfunct1}
\end{align}
where $\,\,r_{aA\,<} = {\rm  min}\,\{r_{aA},r'_{aA}\}$ and $\,\,r_{aA\,>}= {\rm max}\,\{r_{aA},r'_{aA} \}$. The Coulomb regular solution $\varphi_{l_{B}}^{C}(k_{aA},\,r_{aA})$ of the partial Schr\"odinger equation at real momentum $k_{aA}$ is
\begin{align}
&\varphi_{l_{B}}^{C}(k_{aA},\,r_{aA}) =  \frac{1}{2\,i\,k_{aA}}\,\Big[L_{l_{B}}^{C(-)}(k_{aA})\,f_{l_{B}}^{C(+)}(k_{aA},\,r_{aA}) - L_{l_{B}}^{C(+)}(k_{aA})\,f_{l_{B}}^{C(-)}(k_{aA},\,r_{aA})\Big]                                                          \nonumber\\
&= r_{aA}^{l_{B}+1}\,e^{i\,k_{aA}\,r_{aA}}\,{}_1F_{1}(l_{B} + 1 + i\eta_{aA}, 2\,l_{B} + 2; -2\,i\,k_{aA}\,r_{aA})                                    \nonumber\\
&= e^{-i\,\pi\,l_{B}/2}\,L_{l_{B}}^{C(+)}(k_{aA})\,\frac{e^{i\,\sigma_{l_{B}}^{C}  }\,F_{l_{B}}(k_{aA},\,r_{aA})}{k_{aA}},                                                 
\label{varphiFl1}
\end{align}
where
\begin{align}
e^{i\,\sigma_{l_{B}}^{C} }\,F_{l_{B}}(k_{aA},r_{aA})=e^{-\pi\,\eta_{aA}/2}\,\frac{\Gamma(l_{B} + 1 + \,i\,\eta_{aA})}{2\,\Gamma(2\,l_{B} + 2)}\,(2\,k_{aA}\,r_{aA})^{l_{B} + 1}\,e^{i\,k_{aA}\,r_{aA}}\,{}_1F_{1}(l_{B} + 1 + i\,\eta_{aA}, 2\,l_{B} + 2; \, -i\,2\,k_{aA}\,r_{aA}),
\label{Coulregfunct1}
\end{align}
$\sigma_{l_{B}}^{C} $ is the  Coulomb scattering phase shift.
Also
\begin{align}
f_{l_{B}}^{C(\pm)}(k_{aA},r_{aA}) = e^{\pi\,\eta_{aA}/2}\,W_{\mp i\,\eta_{aA},\,l_{B} + 1/2}(\mp 2\,ik_{aA}\,r_{aA})
\label{iostsolution1}
\end{align}
are the Jost solutions (singular at the origin $r_{aA}=0$), 
\begin{align}
L_{l_{B}}^{C(\pm)}(k_{aA}) = \frac{1}{(2\,k_{aA})^{l_{B}}}\,e^{\pi\,\eta_{aA}/2}\,e^{\pm i\,\pi\,l_{B}/2}\,\frac{\Gamma(2\,l_{B} + 2)}{\Gamma(l_{B} +1 \pm i\,\eta_{aA})}
\label{iostfunction1}
\end{align}
are the Jost functions.

Now it is convenient to introduce  the modified Coulomb  wave function
\begin{align}
{\tilde {\varphi}}_{l_{B}}^{C}(k_{aA},\,r_{aA})   =\frac{\varphi_{l_{B}}^{C}(k_{aA},\,r_{aA}) }{L_{l_{B}}^{C(+)}(k_{aA})  },
\label{tilde varphi}
\end{align}
which will be used from now on instead of $\varphi_{l_{B}}^{C}(k_{aA},\,r_{aA})$.

The asymptotic behavior of the overlap function in  (\ref{radeqst2})  is correct because it is governed by the Green function:
\begin{align}
&I_{aA\,\,{l_B}\,{j_B}\,J_{B}}(k_{aA(0)},\,r_{aA}) \stackrel{r_{aA} > R_{aA}}{=} \,-\,2\,\mu_{aA}\,\frac{W_{-i\,\eta_{aA}^{(0)},\,l_{B} + 1/2}(-2\,i\,k_{aA(0)}\,r_{aA})}{R_{aA}}\,e^{\pi\,\eta_{aA}^{(0)}/2}                                                                 \nonumber\\
& \times \int\limits_{0}^{R_{aA}}\,{\rm d}r'_{aA}\,r'_{aA}\,{\tilde \varphi}_{l_{B}}^{C}(k_{aA(0)},\,r_{aA}')\,Q_{l_{B}j_{B}J_{a}J_{A}J_{B}}(r'_{aA}).
\label{asymradovint1}
\end{align}

Replacing the left-hand-side of this equation by Eq. (\ref{OvW1}) one gets the expression of the ANC or the resonance width in terms of the source term:
\begin{align}
{\tilde C}^{B}_{aA\,l_{B}\,j_{B}\,J_{B}}= i^{-l_{B}}\,\,e^{-i\,\delta_{l_{B}}^{hs}}\sqrt{\frac{\mu_{aA}}{k_{aA(0)}}\,\Gamma_{aA\,l_{B}\,j_{B}\,J_{B}}} = -2\,\frac{\mu_{aA}}{k_{aA(0)}}\,i^{-l_{B}}\,\int\limits_{0}^{R_{aA}}\,{\rm d}r'_{aA}\,r'_{aA}\,F_{l_{B}}(k_{aA(0)},\,r_{aA}')\,Q_{l_{B}j_{B}J_{a}J_{A}J_{B}}(r'_{aA}).
\label{asIl2}
\end{align}

This equation provides the ANC or resonance width of the narrow resonance, which may depend on the channel radius $R_{aA}$. Here I am interested in the ratio of the ANC of the resonance state and the ANC of the mirror bound state.   Below will be checked the sensitivity of this ratio to the variation  of the channel radius.

\subsection{ANC in terms of Wronskian}
\label{wronskian}

The advantage of Eq. (\ref{asIl2}) is that to calculate the ANC one needs to know the overlap function only in the nuclear interior where the {\it ab initio} methods like the no-core-shell-model \cite{navratil2000,navratil2003,quaglioni}, and the coupled-cluster method \cite{jensen}  are more accurate than in the external region. Now we transform the radial integral in Eq. (\ref{asIl2}) into the Wronskian at $r_{aA}= R_{aA}$. The philosophy of this transformation is the same as in the surface integral formalism \cite{muk2011,muk2012}.

First, let us rewrite
\begin{align}
V_{aA}- U_{aA}^{C}= V + V_{l_{B}}^{centr}- V_{a} - V_{A} - V_{l_{B}}^{centr} - U_{aA}^{C}
\label{pottransf1}
\end{align} 
and take into account equations 
\begin{align}
(E_{aA(0)} - {\hat T}_{a} - {\hat T}_{A} - {\hat T}_{r_{aA}})\,{\tilde \varphi}_{l_{B}}^{C}(k_{aA(0)},\,r_{aA})\,\varphi_{a}(\xi_{a})\,\varphi_{A}(\xi_{A})= (U_{aA}^{C} + V_{l_{B}}^{centr} + V_{a} + V_{A})\,{\tilde \varphi}_{l_{B}}^{C}(k_{aA(0)},\,r_{aA})\,{\tilde \varphi}_{a}(\xi_{a})\,\varphi_{A}(\xi_{A})
\label{Shreq2}
\end{align}
and 
\begin{align}
(E_{aA(0)} - {\hat T}_{a} - {\hat T}_{A} - {\hat T}_{r_{aA}})\,<Y_{{l_B}{m_{{l_B}}}}({\rm {\bf {\widehat r}}}_{aA})|\Psi(\xi_{a},\,\xi_{A};\,{\rm {\bf r}}_{aA}>\,=\,(V_{aA} + V_{a} + V_{A} + V_{l_{B}}^{centr})\,<Y_{{l_B}{m_{{l_B}}}}({\rm {\bf {\widehat r}}}_{aA})|\Psi(\xi_{a},\,\xi_{A};\,{\rm {\bf r}}_{aA}>,  
\label{eqvarphiB1}
\end{align}
where ${\hat T}_{r_{aA}}$ is the radial kinetic energy operator.

Then we get 
\begin{align}
&{\tilde C}_{aA\,\,{l_B}\,{j_B}\,J_{B}}^{B} \approx -\,2\,\mu_{aA}\,\int\limits_{0}^{R_{aA}}\,{\rm d}r_{aA}\,r_{aA}\,               
{\tilde \varphi}_{l_{B}}^{C}(k_{aA(0)},\,r_{aA})\,Q_{l_{B}j_{B}J_{a}J_
{A}J_{B}}(r_{aA})\,=  \,-\,2\,\mu_{aA}\,                                                         \nonumber\\
& \times\,\sum\limits_{{m_{j_B}{m_{{l_B}}}M_{A}M_{a}}} { < {J_A}{M_A}\,\,{j_B}{m_{{j_B}}}|{J_B}{M_B} >  < {J_a}{M_a}\,{l_B}{m_{{l_B}}}|{j_B}{m_{{j_B}}} > }\,{\left( \begin{gathered}A \hfill \\
a \hfill \\ 
\end{gathered}  \right)^{1/2}}\,\int\limits_{0}^{R_{aA}}\,{\rm d}r_{aA}\,r_{aA}\,
{\tilde \varphi}_{l_{B}}^{C}(k_{aA(0)},\,r_{aA})                  \nonumber\\
&\times\,\int\,{\rm d}\,\Omega_{{\rm {\bf r}}_{r_{aA}}} \,<\varphi_{a}(\xi_{a})\,\varphi_{A}(\xi_{A})|{\overleftarrow {\hat T}}_{r_{aA}} + {\overleftarrow {\hat T}}_{a} + {\overleftarrow {\hat T}}_{A} - {\overrightarrow {\hat T}}_{a} - {\overrightarrow {\hat T}}_{A} - {\overrightarrow {\hat T}}_{r_{aA}}|Y_{{l_B}{m_{{l_B}}}}^{*}({\rm {\bf {\widehat r}}}_{aA})\,\Psi(\xi_{a},\xi_{A};{\rm {\bf r}}_{aA})>                                                                       \nonumber\\
&= -\,2\,\mu_{aA}\,\sum\limits_{{m_{j_B}{m_{{l_B}}}M_{A}M_{a}}} { < {J_A}{M_A}\,\,{j_B}{m_{{j_B}}}|{J_B}{M_B} >  < {J_a}{M_a}\,{l_B}{m_{{l_B}}}|{j_B}{m_{{j_B}}} > }\nonumber\\
& \times\,{\left( \begin{gathered}A \hfill \\
a \hfill \\ 
\end{gathered}  \right)^{1/2}}\,\int\limits_{0}^{R_{aA}}\,{\rm d}r_{aA}\,r_{aA}\,
{\tilde \varphi}_{l_{B}}^{C}(k_{aA(0)},\,r_{aA})\,\int\,{\rm d}\,\Omega_{{\rm {\bf r}}_{aA}}\,<\varphi_{a}(\xi_{a})\,\varphi_{A}(\xi_{A})|{\overleftarrow {\hat T}}_{r_{aA}} - {\overrightarrow {\hat T}}_{r_{aA}}|Y_{{l_B}{m_{{l_B}}}}^{*}({\rm {\bf {\widehat r}}}_{aA})\Psi(\xi_{a},\xi_{A};{\rm {\bf r}}_{aA})>
\nonumber\\
&= -\,2\,\mu_{aA}\,\int\limits_{0}^{R_{aA}}\,{\rm d}r_{aA}\,r_{aA}\,{\tilde \varphi}_{l_{B}}^{C}(k_{aA(0)},\,r_{aA})\,\Big({\overleftarrow {\hat T}}_{r_{aA}} - {\overrightarrow {\hat T}}_{r_{aA}} \Big)\,I_{aA\,\,{l_B}\,{j_B}\,J_{B}}(k_{aA(0)},\,{r_{aA}}).
\label{ANCwronskian1}
\end{align}

Taking into account that 
\begin{align}
f(x)\,\Big(\frac{{\overleftarrow d}^{2}}{dx^{2}}    - \frac{{\overrightarrow d}^{2}}{dx^{2}} \Big)\,g(x\, 
= \frac{d}{dx}\,\Big(g(x)\,\frac{df(x)}{dx} - f(x)\,\frac{dg(x)}{dx} \Big)
\label{greentheorem1}
\end{align}
we arrive at the final expression for the ANC of the resonance state in terms of the Wronskian:
\begin{align}
{\tilde C}_{aA\,\,{l_B}\,{j_B}\,J_{B}}^{B}= \,W[I_{aA\,\,{l_B}\,{j_B}\,J_{B}}(k_{aA(0)},\,{r_{aA}}),\,{\tilde \varphi}_{l_{B}}^{C}(k_{aA(0)},\,r_{aA})]\Big|_{r_{aA}=R_{aA}},
\label{ANCwronskian2}
\end{align}
where the Wronskian 
\begin{align}
&W[\,I_{aA\,\,{l_B}\,{j_B}\,J_{B}}({r_{aA}}),\,{\tilde \varphi}_{l_{B}}^{C}(k_{aA(0)},\,r_{aA})]                                           \nonumber\\
&= I_{aA\,\,{l_B}\,{j_B}\,J_{B}}(k_{aA(0)},\,{r_{aA}})\,\frac{{\rm d} {\tilde \varphi}_{l_{B}}^{C}(k_{aA(0)},\,r_{aA})\,}{{\rm d}r_{aA}}   - {\tilde  \varphi}_{l_{B}}^{C}(k_{aA(0)},\,r_{aA})\,\,\frac{{\rm d}I_{aA\,\,{l_B}\,{j_B}\,J_{B}}(k_{aA(0)},\,{r_{aA}})}{{\rm d}r_{aA}} .
\label{wronskian1}
\end{align} 

We know that the Wronskian calculated for two independent solutions of the Schr\"odinger equation is a constant \cite{newton}. Because the radial overlap function $ I_{aA\,\,{l_B}\,{j_B}\,J_{B}}(k_{aA(0)},\,{r_{aA}})$ is not a solution of the Schr\"odinger equation in the nuclear interior, the Wronskian and, hence, the ANC determined by Eq. (\ref{ANCwronskian2}) depend on the channel radius $R_{aA}$, if it is not too large. However, if the adopted channel radius is large enough, we can replace the radial overlap function by its asymptotic term, see Eq. (\ref{OvW1}), proportional to the Whittaker function, which determines the radial shape of the asymptotic radial overlap function. This Whittaker function is a singular solution of the radial Schr\"odinger equation. \ ${\tilde \varphi}_{l_{B}}^{C}(k_{aA(0)},\,r_{aA})$ is an independent regular solution of the same equation.  Taking into account that $W[\,f_{l_{B}}^{C(+)}(k_{aA(0)},\,r_{aA}),\,f_{l_{B}}^{C(-)}(k_{aA(0)},r_{aA})]= -2\,i\,k_{aA(0)}$ and Eq. (\ref{varphiFl1}) we get at large $R_{aA}$
\begin{align}
W[I_{aA\,\,{l_B}\,{j_B}\,J_{B}}(k_{aA(0)},\,{r_{aA}}),\,{\tilde \varphi}_{l_{B}}^{C}(k_{aA(0)},\,r_{aA})]\Big|_{r_{aA}=R_{aA}}
={\tilde C}_{aA\,\,{l_B}\,{j_B}\,J_{B}}^{B}.
\label{wronskianANC1}
\end{align}  
Hence Eq. (\ref{ANCwronskian2}) at large $R_{aA}$, as expected, turns into identity and the proof of it is an additional test that Eq. (\ref{ANCwronskian2}) is correct. However my idea is to use Eq. (\ref{ANCwronskian2}) at $R_{aA}$, which doesn't
exceed the radius of nucleus $B=(aA)$. In the nuclear interior the contemporary microscopic models can provide quite accurate overlap functions.
The ANC calculated using Eq. (\ref{ANCwronskian2}) may depend on the adopted channel radius $R_{aA}$ but the sensitivity to the variation of the channel radius of the ratio of the ANCs of the resonance and mirror bound state is significantly  weaker than that of the individual ANCs (or, equivalently, of the resonance width and the ANC) of the mirror states. 
This ratio of the ANCs  of the  resonance $B_{1}=(a_{1}\,A_{1})$ and the mirror bound state $B_{2}=(a_{2}\,A_{2})$ is given by:
\begin{align}
L^{W}(R_{ch}) = \frac{\Big|({\tilde C}_{a_{1}A_{1}\,\,{l_B}\,{j_B}\,J_{B}}^{B_{1}})\Big|^{2}}{({C}_{a_{2}\,A_{2}\,\,{l_B}\,{j_B}\,J_{B}}^{B_{2}})^{2}}\, =\, 
\frac{\Big|W[I_{a_{1}A_{1}\,\,{l_B}\,{j_B}\,J_{B}}(k_{a_{1}A_{1}(0)},\,{r_{a_{1}A_{1}}}),\,{\tilde \varphi}_{l_{B}}^{C}(k_{a_{1}A_{1}(0)},\,r_{a_{1}A_{1}})]\Big|^{2}\,\Big|_{r_{a_{1}A_{1}}=R_{ch}}}{\Big(W[I_{a_{2}A_{2}\,\,{l_B}\,{j_B}\,J_{B}}(\kappa_{a_{2}A_{2}},\,r_{a_{2}A_{2}} ),\,{\tilde \varphi}_{l_{B}}^{C}(\kappa_{a_{2}A_{2}},\,r_{a_{2}A_{2}})]\Big)^{2}\,\Big|_{r_{a_{2}A_{2}}=R_{ch}}}.
\label{ratioANCmirrorratio1}
\end{align}
 Because $B_{1}$ and $B_{2}$ are mirror nuclei, the quantum numbers in both nuclei are the same. We also assume that the channel radius $R_{ch}$ is the same for both mirror nuclei.
 
 Taking into account Eq. (\ref{asIl2}) we get for the ratio of the resonance width and the bound state ANC for mirror states:
 \begin{align}
L_{1}^{W}(R_{ch}) = \frac{ \Gamma_{a_{1}A_{1}\,\,{l_B}\,{j_B}\,J_{B}}}{({C}_{a_{2}\,A_{2}\,\,{l_B}\,{j_B}\,J_{B}}^{B_{2}})^{2}}\, =\,\sqrt{\frac{2\,E_{a_{1}A_{1}(0)}}{\mu_{a_{1}A_{1}}}}\,
\frac{\Big|W[I_{a_{1}A_{1}\,\,{l_B}\,{j_B}\,J_{B}}(k_{a_{1}A_{1}(0)},\,{r_{a_{1}A_{1} } }),\,{\tilde \varphi}_{l_{B}}^{C}(k_{a_{1}A_{1}(0)},\,r_{a_{1}A_{1}})]\Big|^{2}\,\Big|_{r_{a_{1}A_{1}}=R_{ch}}}{\Big(W[I_{a_{2}A_{2}\,\,{l_B}\,{j_B}\,J_{B}}(\kappa_{a_{2}A_{2}},\,r_{a_{2}A_{2}} ),\,{\tilde \varphi}_{l_{B}}^{C}(\kappa_{a_{2}A_{2}},\,r_{a_{2}A_{2}})]\Big)^{2}\,\Big|_{r_{a_{2}A_{2}}=R_{ch}}},
\label{GammaANCmirrorratio1}
\end{align}
where $E_{a_{1}A_{1}(0)}$ and $\mu_{aA}$ are expressed in MeV.
 Thus the ratio of the resonance width and  the square of the ANC of the mirror state is expressed in terms of the ratio of the resonant and bound states Wronskians. Equation (\ref{GammaANCmirrorratio1}) allows one to determine the resonance width if the  ANC of the mirror bound state is known and vice versa.  
  
To calculate  the ratio   $\frac{\Gamma_{a_{1}A_{1}\,\,{l_B}\,{j_B}\,J_{B}}^{B_{1}}}{({C}_{a_{2}\,A_{2}\,\,{l_B}\,{j_B}\,J_{B}}^{B_{2}})^{2}}$ 
 one needs the microscopic radial overlap functions.  If  these radial overlap functions are not available then one  can use a standard  approximation for the overlap functions:
 \begin{align}
 &I_{a_{1}A_{1}\,\,{l_B}\,{j_B}\,J_{B}}(k_{a_{1}A_{1}(0)},\,{r_{aA}}) \approx  S_{a_{1}A_{1}}^{1/2}\, \varphi_{a_{1}A_{1}\,\,{l_B}\,{j_B}\,J_{B}}(k_{a_{1}A_{1}(0)},\,{r_{a_{1}A_{1}}}),\\
& I_{a_{2}A_{2}\,\,{l_B}\,{j_B}\,J_{B}}(\kappa_{a_{2}A_{2}},\,{r_{aA}}) \approx  S_{a_{2}A_{2}}^{1/2}\, \varphi_{a_{2}A_{2}\,\,{l_B}\,{j_B}\,J_{B}}(\kappa_{a_{2}A_{2}},\,{r_{a_{2}A_{2}}}),
 \label{approxOFvarphi1}
 \end{align}
 where $S_{a_{1}A_{1}}$ and $S_{a_{2}A_{2}}$ are the spectroscopic factors of the mirror resonance and bound states $(a_{1}A_{1})$ and $(a_{2}A_{2})$, respectively.
  $\varphi_{a_{1}A_{1}\,\,{l_B}\,{j_B}\,J_{B}}(k_{a_{1}A_{1}(0)},\,{r_{a_{1}A_{1}}})$ is a real internal resonant wave  function calculated in the two-body model 
  $(a_{1} \, A_{1})$ using some phenomenological potential, for example, Woods-Saxon one, which supports the resonance state under consideration.    $\varphi_{a_{2}A_{2}\,\,{l_B}\,{j_B}\,J_{B}}(\kappa_{a_{2}A_{2}},\,{r_{a_{2}A_{2}}})$  is the two-body bound-state wave function of the bound state $(a_{2}\,A_{2})$, which is also calculated using the same nuclear potential as the mirror resonance state.  If the mirror symmetry holds then $S_{a_{1}A_{1}} \approx S_{a_{2}A_{2}}$ and we get
an approximated   $\frac{ \Gamma_{a_{1}A_{1}\,\,{l_B}\,{j_B}\,J_{B}}}{({C}_{a_{2}\,A_{2}\,\,{l_B}\,{j_B}\,J_{B}}^{B_{2}})^{2}}$  ratio in terms of the Wronskians, which does not contain the overlap functions:
\begin{align}
L_{2}^{W}(R_{ch}) =\frac{ \Gamma_{a_{1}A_{1}\,\,{l_B}\,{j_B}\,J_{B}}}{({C}_{a_{2}\,A_{2}\,\,{l_B}\,{j_B}\,J_{B}}^{B_{2}})^{2}}\, =\,\sqrt{\frac{2\,E_{a_{1}A_{1}(0)}}{\mu_{a_{1}A_{1}}}}\,
\frac{\Big|W[\varphi_{a_{1}A_{1}\,\,{l_B}\,{j_B}\,J_{B}}(k_{a_{1}A_{1}(0)},\,{r_{a_{1}A_{1} } }),\,{\tilde \varphi}_{l_{B}}^{C}(k_{a_{1}A_{1}(0)},\,r_{a_{1}A_{1}})]\Big|^{2}\,\Big|_{r_{a_{1}A_{1}}=R_{ch}}}{\Big(W[\varphi_{a_{2}A_{2}\,\,{l_B}\,{j_B}\,J_{B}}(\kappa_{a_{2}A_{2}},\,r_{a_{2}A_{2}} ),\,{\tilde \varphi}_{l_{B}}^{C}(\kappa_{a_{2}A_{2}},\,r_{a_{2}A_{2}})]\Big)^{2}\,\Big|_{r_{a_{2}A_{2}}=R_{ch}}}.
\label{GammaANCmirrorratio2}
\end{align}
 
 Meantime in \cite{timofeyuk2003} another expression for the mirror nucleon ANCs ratio was obtained which provides the easiest way to determine  $\frac{ \Gamma_{a_{1}A_{1}\,\,{l_B}\,{j_B}\,J_{B}}}{({C}_{a_{2}\,A_{2}\,\,{l_B}\,{j_B}\,J_{B}}^{B_{2}})^{2}}$. I will show here a simple way of the derivation of the ratio  $\frac{ \Gamma_{a_{1}A_{1}\,\,{l_B}\,{j_B}\,J_{B}}}{({C}_{a_{2}\,A_{2}\,\,{l_B}\,{j_B}\,J_{B}}^{B_{2}})^{2}}$   from \cite{timofeyuk2003}.
First, as it was pointed out in \cite{timofeyuk2003}, in the nuclear interior the Coulomb interaction varies very little in the vicinity of $R_{ch}$ and its effect leads only to shifting of the  binding energy of the bound state to the continuum.  Hence, it can be assumed that $\tilde \varphi _{{l_B}}^C({k_{{a_1}{A_1}(0)}},\,{r_{{a_1}{A_1}}})$  and ${\tilde \varphi}_{l_{B}}^{C}(\kappa_{a_{2}A_{2}}\,r_{a_{2}A_{2}})$ behave similarly near $ R_{ch}$ except for the overall normalization, that is 
\begin{align}
\tilde \varphi _{{l_B}}^C({k_{{a_1}{A_1}(0)}},\,{r_{aA}}) = \frac{{\tilde \varphi _{{l_B}}^C({k_{{a_1}{A_1}(0)}},{R_{ch}})}}{{\tilde \varphi _{{l_B}}^C(\,{\kappa _{{a_2}{A_2}}},\,{R_{ch}})\,}}\,\tilde \varphi _{{l_B}}^C({\kappa _{{a_2}{A_2}}},\,{r_{{a_2}{A_2}}}).
\label{varphisimple1}
\end{align}

Then
\begin{align}
& L_{3}^{W}(R_{ch}) =\frac{ \Gamma_{a_{1}A_{1}\,\,{l_B}\,{j_B}\,J_{B}}}{({C}_{a_{2}\,A_{2}\,\,{l_B}\,{j_B}\,J_{B}}^{B_{2}})^{2}} = \sqrt{\frac{{2\,{E_{{a_1}{A_1}(0)}}}}{{{\mu _{{a_1}{A_1}}}}}}\,{\Big(\frac{{\tilde \varphi _{{l_B}}^C({k_{{a_1}{A_1}(0)}},{R_{ch}}}}{{\tilde \varphi _{{l_B}}^C({\kappa _{{a_2}{A_2}}},\,{R_{ch}}){\rm{ }}}}\Big)^2}
     \nonumber\\
& \times \frac{{|W[{\varphi_{{a_1}{A_1}\,\,{l_B}\,{j_B}\,{J_B}}}({k_{{a_1}{A_1}(0)}},\,{r_{{a_1}{A_1}}}),\,\tilde \varphi _{{l_B}}^C(i\,{\kappa _{{a_2}{A_2}}},{r_{{a_2}{A_2}}})]\,{|^2}\,{|_{{r_{{a_1}{A_1}}} = {R_{ch}}}}}}{{{{(W[{\varphi_{{a_2}{A_2}\,\,{l_B}\,{j_B}\,{J_B}}}({\kappa _{{a_2}{A_2}}},\,{r_{{a_2}{A_2}}}),\,\tilde \varphi _{{l_B}}^C({\kappa _{{a_2}{A_2}}},\,{r_{{a_2}{A_2}}})])}^2}\,{|_{{r_{{a_2}{A_2}}} = {R_{ch}}}}}}.
\label{Ratioapprox1}
\end{align}

Neglecting further the difference between the mirror wave functions ${\varphi _{{a_1}{A_1}\,\,{l_B}\,{j_B}\,{J_B}}}({k_{{a_1}{A_1}(0)}},\,{r_{{a_1}{A_1}}})$ and ${\varphi_{{a_2}{A_2}\,\,{l_B}\,{j_B}\,{J_B}}}({\kappa _{{a_2}{A_2}}},\,{r_{{a_2}{A_2}}})$  in the nuclear interior we obtain the approximate expression 
for $\frac{ \Gamma_{a_{1}A_{1}\,\,{l_B}\,{j_B}\,J_{B}}}{({C}_{a_{2}\,A_{2}\,\,{l_B}\,{j_B}\,J_{B}}^{B_{2}})^{2}}\,$  from \cite{timofeyuk2003} (in the notations of the current paper):
\begin{align}
&{L_{4}^{W}}({R_{ch}}) = \frac{ \Gamma_{a_{1}A_{1}\,\,{l_B}\,{j_B}\,J_{B}}}{({C}_{a_{2}\,A_{2}\,\,{l_B}\,{j_B}\,J_{B}}^{B_{2}})^{2}}= \sqrt{\frac{{2\,{E_{{a_1}{A_1}(0)}}}}{{{\mu _{{a_1}{A_1}}}}}}\,{\Big(\frac{{\tilde \varphi _{{l_B}}^C(k_{{a_1}{A_1}(0)}^{{B_1}},\,{R_{ch}})}}{{\tilde \varphi _{{l_B}}^C(i\,{\kappa _{{a_2}{A_2}}},\,{R_{ch}})}}\Big)^2}.
\label{Ratioapprox2}
\end{align} 

In descending accuracy I can rank Eq. (\ref{GammaANCmirrorratio1}) as the most accurate, then Eq. (\ref{Ratioapprox1}). Taking into account that the microscopic overlap functions (calculated in the no-core-shell-model \cite{navratil2000,navratil2003,quaglioni} or oscillator shell-model \cite{tim2011}) are  accurate in the nuclear interior, using Eq (\ref{GammaANCmirrorratio1}) one can determine the ratio  $\frac{ \Gamma_{a_{1}A_{1}\,\,{l_B}\,{j_B}\,J_{B}}}{({C}_{a_{2}\,A_{2}\,\,{l_B}\,{j_B}\,J_{B}}^{B_{2}})^{2}}$  quite accurately.   Then follows  Eq. (\ref{GammaANCmirrorratio2}) and finally  Eq. (\ref{Ratioapprox2}).  
Note that Eq. (\ref{Ratioapprox2})  is valid  only in the region where the mirror resonant and bound state wave functions do coincide or very close.  The advantage of this  equation is that it allows one to calculate the ratio without using the mirror wave functions and extremely simple to use.

Because for the cases under consideration   the internal microscopic resonance wave functions are not available,  in this paper  the    $\frac{ \Gamma_{a_{1}A_{1}\,\,{l_B}\,{j_B}\,J_{B}}}{({C}_{a_{2}\,A_{2}\,\,{l_B}\,{j_B}\,J_{B}}^{B_{2}})^{2}}$ ratio is calculated  using  Eqs (\ref{GammaANCmirrorratio2}) and (\ref{Ratioapprox2}).
It allows one to determine the accuracy of both equations. 

Note that the dimension of the ratio $\frac{ \Gamma_{a_{1}A_{1}\,\,{l_B}\,{j_B}\,J_{B}}}{({C}_{a_{2}\,A_{2}\,\,{l_B}\,{j_B}\,J_{B}}^{B_{2}})^{2}}$ is determined by the ratio
$ \frac{{2\,{E_{{a_1}{A_1}(0)}}}}{{{\mu _{{a_1}{A_1}}}}}$. To make it dimensionless I assume that the reduced mass  $\mu _{{a_1}{A_1}}$ and the real part of the resonance energy  $E_{{a_1}{A_1}(0)}$ are expressed in MeV.

\section{Comparison  of  resonance widths and  ANCs of  mirror states}

In this section a few examples of  the application of  Eqs. (\ref{GammaANCmirrorratio2}) and (\ref{Ratioapprox2}) are presented. To simplify the notations from now on  the quantum numbers in the notations for the resonance width and the ANC are dropped and just use simplified notations,
$\Gamma_{a_{1}A_{1}}$ and ${C}_{a_{2}\,A_{2}}$.  
Equation  (\ref{GammaANCmirrorratio2}) gives  $\Gamma_{a_{1}A_{1}}/({C}_{a_{2}\,A_{2}})^{2}$ in terms of the ratio of the Wronskians and 
 provides an exact value  for given two-body mirror resonant and bound-state wave functions.  Equation  (\ref{Ratioapprox2})   gives  the $\Gamma_{a_{1}A_{1}}/({C}_{a_{2}\,A_{2}})^{2}$   ratio in terms of the  Coulomb scattering wave functions at the real resonance momentum $k_{{a_1}{A_1}(0)}$  and the imaginary momentum of the bound state 
 $i\,\kappa _{{a_2}{A_2}}$  at the channel radius $R_{ch}$.  Hence, to determine the ratio  $\Gamma_{a_{1}A_{1}}/({C}_{a_{2}\,A_{2}})^{2}$ using Eq. (\ref{Ratioapprox2})  one does not need to know the mirror  resonant and bound-state wave functions. However, to use this equation one should check whether the mirror wave functions are close.
  
\subsection{Comparison of resonance width for $\mathbf{{}^{13}{\rm N}(2s_{1/2}) \to {}^{12}{\rm C}(0.0\,{\rm MeV}) + p}$ and mirror ANC for virtual decay $\mathbf{{}^{13}{\rm C}(2s_{1/2}) \to {}^{12}{\rm C}(0.0\,{\rm MeV}) + n}$}

I begin from the analysis of the isobaric analogue states  $2s_{1/2}$ in the mirror nuclei ${}^{13}{\rm N}$ and ${}^{13}{\rm C}$. 
The resonance energy of   ${}^{13}{\rm N}(2s_{1/2})$  is  $E_{p{}^{12}{\rm C}(0)} = 0.421$ MeV with the resonance width of $\Gamma_{p{}^{12}{\rm C}}= 0.0317 \pm 0.0008$ MeV \cite{AjzenbergSelove}. The neutron binding energy of the mirror state  ${}^{13}{\rm C}(2s_{1/2})$  is $\varepsilon_{n{}^{12}{\rm C}} = 1.8574$ MeV with the experimental ANC $C_{n{}^{12}{\rm C}}^{2}= 3.65$ fm${}^{-1}$ \cite{liu,imai}.  The experimental ratio ${\Gamma _{{p^{12}}{\rm{C}}}}/(C_{{n^{12}}{\rm{C}}})^2 = (4.40 \pm 0.57) \times {10^{ - 5}}$ allows us to check the accuracy of  both used equations.  Because the dimension of the bound-state ANC is fm$^{-1/2}$ to get the dimensionless ratio I calculated  
${\Gamma _{{p^{12}}{\rm{C}}}}/[\hbar\,c(C_{{n^{12}}{\rm{C}}})^2]$.

\begin{figure}[htbp] 
\includegraphics[width=0.5\textwidth]{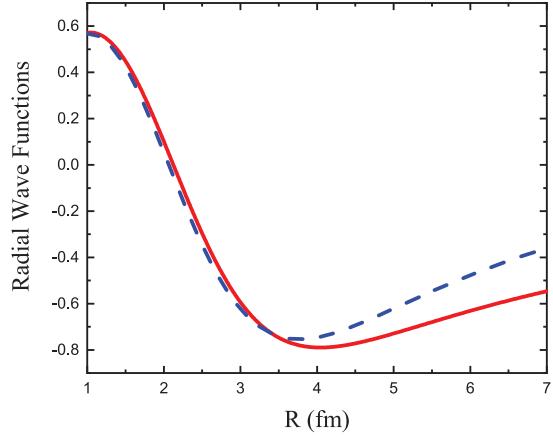}
\caption{ Solid red line:  The  radial wave function of the  $(p{}^{12}{\rm C})_{2s_{1/2}^{+}}$  resonance state; dashed blue line:  the radial wave function of  the mirror    $(n{}^{12}{\rm C})_{2s_{1/2}^{+}}$  bound-state.  $r$ is the distance between $N$, where $N=p,n$, and the c.m. of ${}^{12}{\rm C}$.}
\label{fig_wf13N13C12p1}
\end{figure}

In Fig. \ref{fig_wf13N13C12p1} are shown the radial wave functions of the mirror states.
Following the  $R$-matrix procedure, both  wave functions are normalized to unity over the internal volume with the radius $R_{ch}=4.0$ fm.  We see that the mirror wave functions are very close at distances $\leq 4.0$ fm what confirms the mirror symmetry of  $(p{}^{12}{\rm C})_{2s_{1/2}^{+}}$  and  $(n{}^{12}{\rm C})_{2s_{1/2}^{+}}$ systems. 

\begin{figure}[htbp] 
\includegraphics[width=0.5\textwidth]{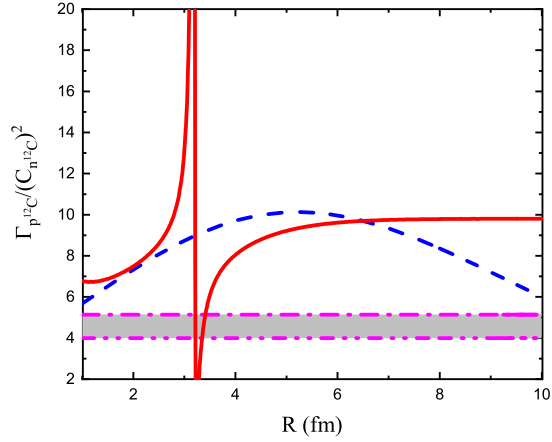}
\caption{ The grey band  is the experimental   $\frac{{{\Gamma _{p{\,^{12}}{\rm{C}}}}}}{{{{({{C}_{n{\,^{12}}{\rm{C}}}})}^2}}}$  ratio of the resonance width of the resonance state ${}^{13}{\rm N}(2s_{1/2}^{+})$ and  the ANC of  the mirror bound state ${}^{13}{\rm C}(2s_{1/2}^{+})$; 
the red dashed-dotted-dotted line and the red dashed-dotted lines are  the low and upper limits of this experimental ratio; 
the solid red line is the  $\frac{{{\Gamma _{p{\,^{12}}{\rm{C}}}}}}{{{{({{C}_{n{\,^{12}}{\rm{C}}}})}^2}}}$  ratio as a function of $R_{ch}$ calculated using Eq. (\ref{GammaANCmirrorratio2});   the blue dotted line  is the  $\frac{{{\Gamma _{p{\,^{12}}{\rm{C}}}}}}{{{{({{C}_{n{\,^{12}}{\rm{C}}}})}^2}}}$ ratio  calculated  as a function of $R_{ch}$ using Eq. (\ref{Ratioapprox2}).}
\label{fig_GANCratio1}
\end{figure}

In  Fig. \ref{fig_GANCratio1}  are shown  the   $\frac{{{\Gamma _{p{\,^{12}}{\rm{C}}}}}}{{{{({{C}_{n{\,^{12}}{\rm{C}}}})}^2}}}$  ratios, which are calculated   
using Eqs (\ref{GammaANCmirrorratio2})   and  (\ref{Ratioapprox2}). These calculated ratios  are  compared with the experimental one. We see that the calculations exceed the experimental value. 
The  $\frac{{{\Gamma _{p{\,^{12}}{\rm{C}}}}}}{{{{({{C}_{n{\,^{12}}{\rm{C}}}})}^2}}}$ ratio calculated using the simplified Eq. (\ref{Ratioapprox2})  shows the $R_{ch}$ dependence  and is equal to $10.13 \times 10^{-5}$ at the peak at $R_{ch}=5.22$ fm.

Equation (\ref{GammaANCmirrorratio2})  provides  the   $\frac{{{\Gamma _{p{\,^{12}}{\rm{C}}}}}}{{{{({{C}_{n{\,^{12}}{\rm{C}}}})}^2}}}$ ratio  in terms of the ratio of the Wronskians.
Each Wronskian  contains  the two-body wave function and its radial derivative of the system  $(N\,{}^{12}{\rm C})_{2s_{1/2}^{+}}$, $N=p,n$. Each two-body wave function has one node at $r \approx 2.13$ fm and a minimum at  $r \approx 4.0$ fm. . Hence, at some point $r$ the Wronskian in the denominator of Eq. (\ref{GammaANCmirrorratio2}) vanishes causing a discontinuity in the ratio  $\frac{{{\Gamma _{p{\,^{12}}{\rm{C}}}}}}{{{{({{C}_{n{\,^{12}}{\rm{C}}}})}^2}}}$.  
 I assume that  in the nuclear interior the mirror two-body wave functions are correct 
(as it should be for the mirror microscopic overlap functions) and calculate the ratio at $E_{ch} \geq 4$ fm.  At  $r=4$ fm  $\frac{{{\Gamma _{p{\,^{12}}{\rm{C}}}}}}{{{{({{C}_{n{\,^{12}}{\rm{C}}}})}^2}}}= 8.1 \times 10^{-5}$ while the correct value of this ratio obtained at large $R_{ch}$  is $9.8 \times 10^{-5}$, which is close to the peak value of the ratio obtained using Eq. (\ref{Ratioapprox2}). 

Both used equations provide the values of the $\frac{{{\Gamma _{p{\,^{12}}{\rm{C}}}}}}{{{{({{C}_{n{\,^{12}}{\rm{C}}}})}^2}}}$ ratio, which exceed the experimental one.
It means that more accurate internal overlap functions are required and  the two-body wave functions  used here demonstrate the accuracy of the Wronskian method.
However, there is another important conclusion:  the simple Eq. (\ref{Ratioapprox2})  in the peak gives the same result  as the asymptotic ratio given by Eq. (\ref{GammaANCmirrorratio2}).

\subsection{Comparison of resonance width for $\mathbf{{}^{13}{\rm N}(1d_{5/2}) \to {}^{12}{\rm C}(0.0\,{\rm MeV}) + p}$ and mirror ANC for virtual decay $\mathbf{{}^{13}{\rm C}(1d_{5/2}) \to {}^{12}{\rm C}(0.0\,{\rm MeV}) + n}$}

As the second example I consider  the isobaric analogue states  $1d_{5/2}$ in the mirror nuclei ${}^{13}{\rm N}$ and ${}^{13}{\rm C}$. 
The resonance energy of   ${}^{13}{\rm N}(1d_{5/2})$  is  $E_{p{}^{12}{\rm C}(0)} =1.6065$ MeV with the resonance width of $\Gamma_{p\,{}^{12}{\rm C}}= 0.047 \pm 0.0008$ MeV \cite{AjzenbergSelove}. The neutron binding energy of the mirror state  ${}^{13}{\rm C}(1d_{5/2})$  is $\varepsilon_{n{}^{12}{\rm C}} = 1.09635$ MeV with the experimental ANC $C_{n{}^{12}{\rm C}}^{2}= 0.0225$ fm${}^{-1}$ \cite{liu}.  The experimental ratio is ${\Gamma _{{p\,^{12}}{\rm{C}}}}/C_{{n\,^{12}}{\rm{C}}}^2 = (1.1 \pm 0.2) \times {10^{ - 2}}$ .

\begin{figure}[htbp] 
\includegraphics[width=0.5\textwidth]{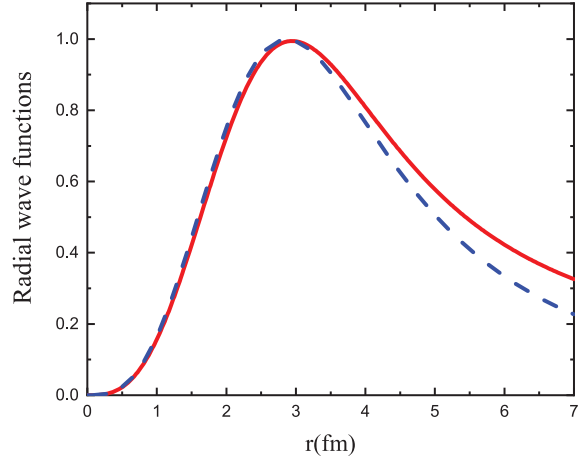}
\caption{ Solid red line:  The  radial wave function of the  $(p{}^{12}{\rm C})_{1d_{5/2}^{+}}$  resonance state; dashed blue line:  the radial wave function of  the mirror    $(n{}^{12}{\rm C})_{1d_{5/2}^{+}}$  bound-state.  $r$ is the distance between $N$, where $N=p,n$, and the c.m. of ${}^{12}{\rm C}$.}
\label{fig_wf13N13C52p1}
\end{figure}

In Fig. \ref{fig_wf13N13C52p1} are shown the radial wave functions of the mirror states.
Following the  $R$-matrix procedure, both  wave functions are normalized to unity over the internal volume with the radius $R_{ch}=3$ fm.  We see that the mirror wave functions are very close at distances $r \leq 4$ fm what confirms the mirror symmetry of  $(p{}^{12}{\rm C})_{1d_{5/2}^{+}}$  and  $(n{}^{12}{\rm C})_{1d_{5/2}^{+}}$ systems. 
\begin{figure}[htbp] 
\includegraphics[width=0.5\textwidth]{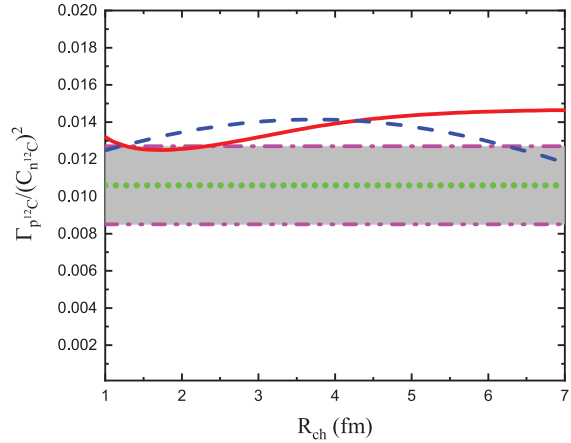}
\caption{The grey band  is the experimental   $\frac{{{\Gamma _{p{\,^{12}}{\rm{C}}}}}}{{{{({{C}_{n{\,^{12}}{\rm{C}}}})}^2}}}$  ratio of the resonance width of the resonance state ${}^{13}{\rm N}(1d_{5/2}^{+})$ and  the ANC of  the mirror bound state ${}^{13}{\rm C}(1d_{5/2}^{+})$; 
the red dashed-dotted-dotted line and the red dashed-dotted lines are  the low and upper limits of this experimental ratio;  the green dotted line is the adopted experimental value of the  
ratio  $\frac{{{\Gamma _{p{\,^{12}}{\rm{C}}}}}}{{{{({{C}_{n{\,^{12}}{\rm{C}}}})}^2}}}= (1.1 \pm 0.2) \times {10^{ - 2}}$ ;
the solid red line is the  $\frac{{{\Gamma _{p{\,^{12}}{\rm{C}}}}}}{{{{({{C}_{n{\,^{12}}{\rm{C}}}})}^2}}}$  ratio as a function of $R_{ch}$ calculated using Eq. (\ref{GammaANCmirrorratio2});   the blue dotted line  is the  $\frac{{{\Gamma _{p{\,^{12}}{\rm{C}}}}}}{{{{({{C}_{n{\,^{12}}{\rm{C}}}})}^2}}}$ ratio  calculated  as a function of $R_{ch}$ using Eq. (\ref{Ratioapprox2}).}
\label{fig_GANC13N13C52p1}
\end{figure}
In  Fig. \ref{fig_GANC13N13C52p1}  are shown  the   $\frac{{{\Gamma _{p{\,^{12}}{\rm{C}}}}}}{{{{({{C}_{n{\,^{12}}{\rm{C}}}})}^2}}}$  ratios  calculated
using Eqs (\ref{GammaANCmirrorratio2})   and  (\ref{Ratioapprox2}), which  are  compared with the experimental ratio. We see that the calculated ratios are closer to the experimental ratio
than in the previous case and both equations give quite reasonable results. 
The  $\frac{{{\Gamma _{p{\,^{12}}{\rm{C}}}}}}{{{{({{C}_{n{\,^{12}}{\rm{C}}}})}^2}}}$ ratio calculated using the simplified Eq. (\ref{Ratioapprox2})  shows the $R_{ch}$ dependence  and is equal to $0.0141$ at the peak at $R_{ch}=3.95$ fm. In the case under consideration the bound-state wave function does not have nodes at $r>0$. That is why 
the  $\frac{{{\Gamma _{p{\,^{12}}{\rm{C}}}}}}{{{{({{C}_{n{\,^{12}}{\rm{C}}}})}^2}}}$ ratio calculated using Eq. (\ref{GammaANCmirrorratio2})  is a smooth function of $R_{ch}$ .
This equation gives  $\frac{{{\Gamma _{p{\,^{12}}{\rm{C}}}}}}{{{{({{C}_{n{\,^{12}}{\rm{C}}}})}^2}}}=0.0135$ at $R_{ch}= 4$ fm, which  differs very little   from its correct asymptotic  value of $0.0143$. Again, as in the previous case, our calculations show that the simple Eq.  (\ref{Ratioapprox2})  can give the results close to the Wronskian method.

\subsection{Comparison of resonance width for $\mathbf{{}^{15}{\rm F}(1d_{5/2}) \to {}^{14}{\rm O}(0.0\,{\rm MeV}) + p}$ and mirror ANC for virtual decay $\mathbf{{}^{15}{\rm C}(1d_{5/2}) \to {}^{14}{\rm C}(0.0\,{\rm MeV}) + n}$}

In this section I determine the ratio ${\Gamma _{{p\,^{14}}{\rm{O}}}}/C_{{n\,^{14}}{\rm{C}}}^2$  for the mirror states  ${}^{15}{\rm F}(1d_{5/2})$ and ${}^{15}{\rm C}(1d_{5/2})$.
The resonance energy and the resonance width of  ${}^{15}{\rm F}(1d_{5/2})$  are  $E_{p{}^{14}{\rm O}(0)}=2.77$ MeV and $\Gamma _{p\,^{14}{\rm O}}= 0.24 \pm 0.03$ MeV   \cite{Tilley}.  
The binding energy and the ANC of the bound state  ${}^{15}{\rm C}(1d_{5/2})$ are $\varepsilon_{n{}^{14}{\rm C}} = 0.478$ MeV and $C_{n{}^{14}{\rm C}}^{2}= (3.6 \pm 0.8) \times 10^{-3}$  fm$^{-1}$.  The experimental ratio ${\Gamma _{{p\,^{14}}{\rm{O}}}}/C_{{n\,^{14}}{\rm{C}}}^2= 0.338 \pm 0.001$ . 

\begin{figure}[htbp] 
\includegraphics[width=0.5\textwidth]{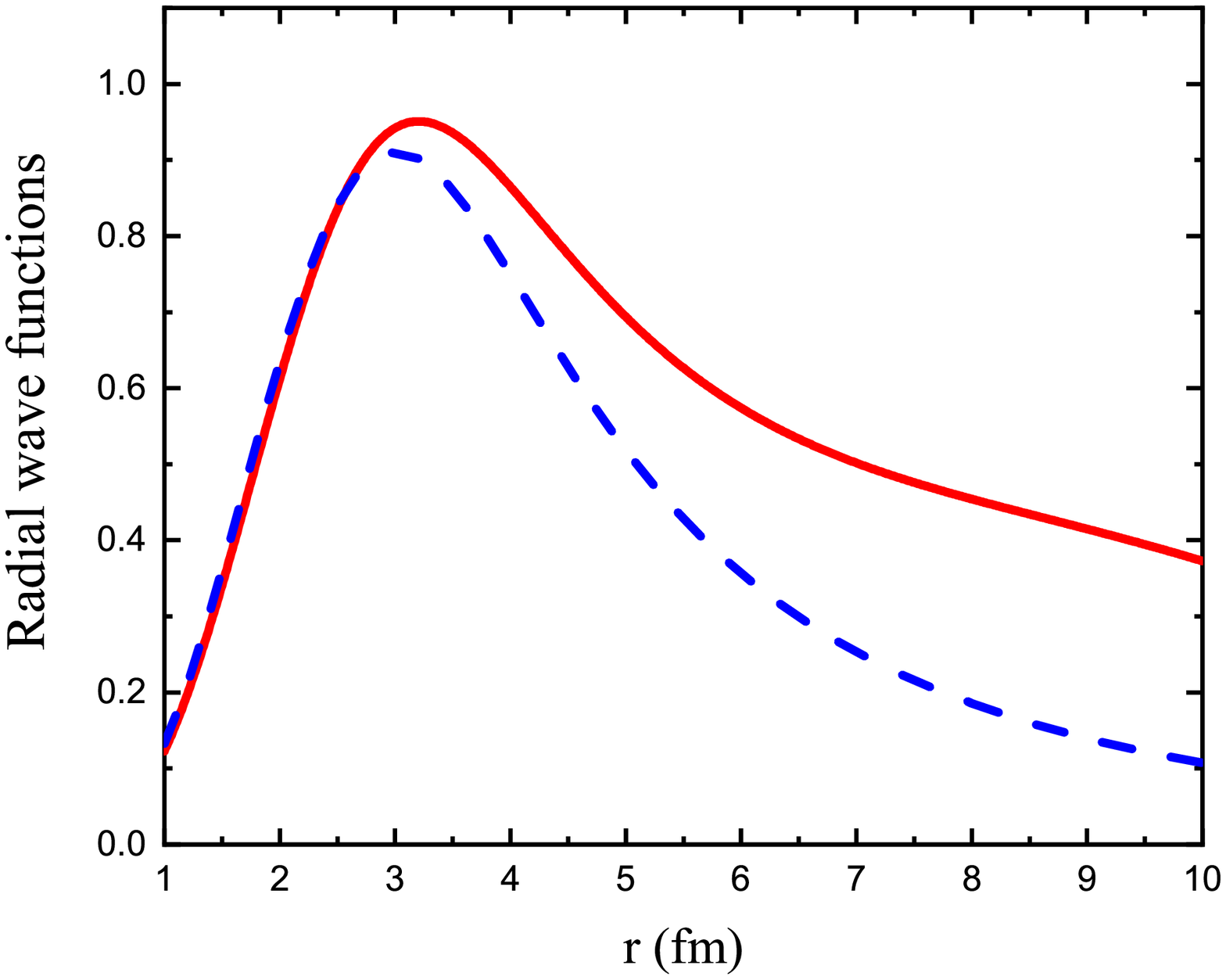}
\caption{ Solid red line:  the  radial wave function of the  $(p{}^{14}{\rm O})_{1d_{5/2}}$  resonance state; dashed blue line:  the radial wave function of  the mirror    $(n{}^{14}{\rm C})_{1d_{5/2}}$  bound-state.  $r$ is the distance between  the nucleon  and the c.m. of the nucleus.}
\label{fig_radwfs15F15C}
\end{figure}

This is the most difficult case because the resonance state is not potential. It is clear from  Fig. \ref{fig_radwfs15F15C}.

The mirror wave functions begin to deviate for  $r> 3.5$ fm.  Because the resonance width in the case under consideration is much wider than in the previous cases, the calculated in the potential model resonant wave function in the external region differs significantly from the tail of the bound-state wave function. That is why the Wronskian ratio does not have an asymptote at large $r$.  But the idea
of the  Wronskian method is to determine the  ${\Gamma _{{p\,^{14}}{\rm{C}}}}/C_{{n\,^{14}}{\rm{C}}}^2$ ratio using the mirror wave functions in the internal region where they   practically coincide.

\begin{figure}[htbp] 
\includegraphics[width=0.5\textwidth]{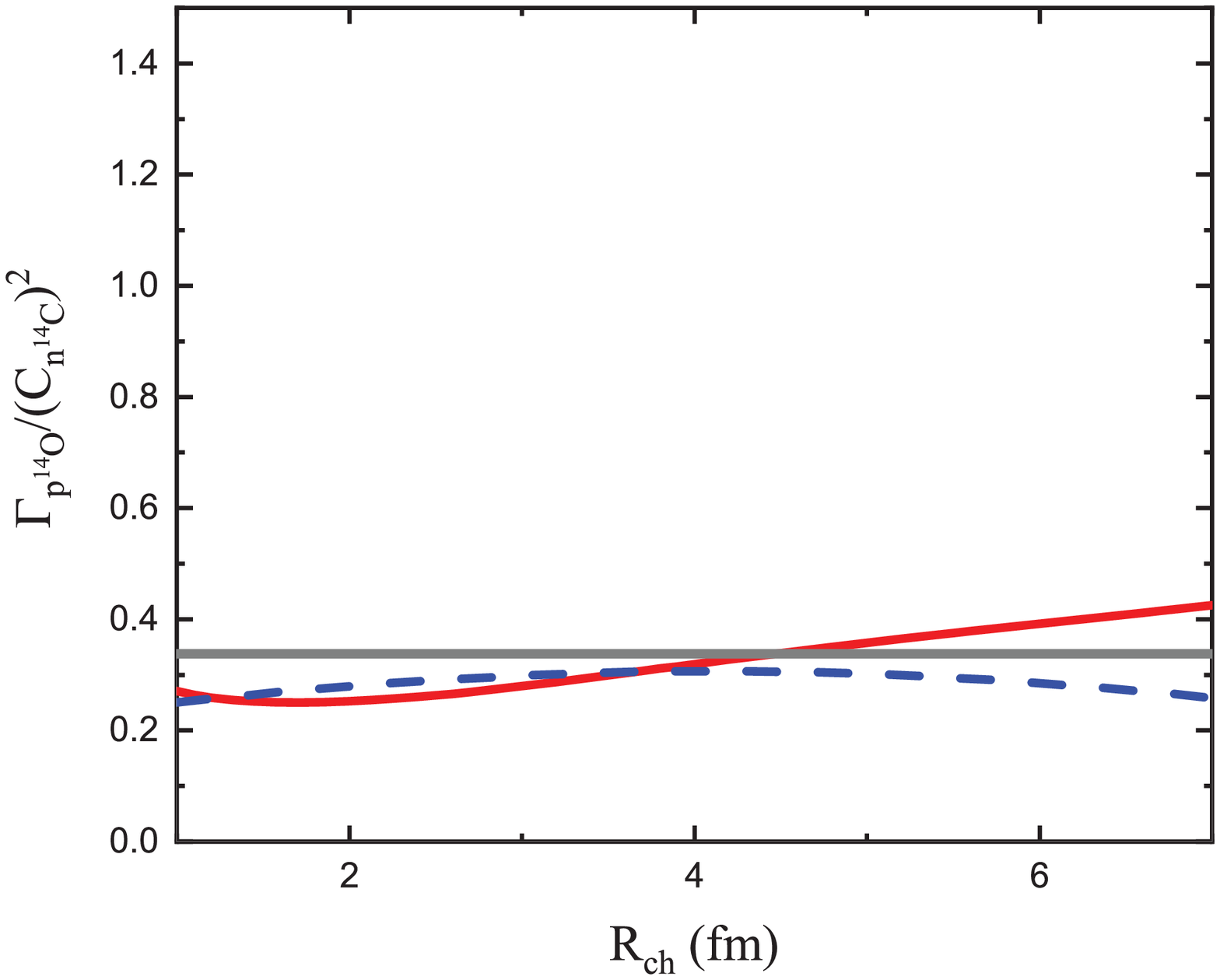}
\caption{ The grey band  is the experimental   $\frac{{{\Gamma _{p{\,^{14}}{\rm{O}}}}}}{{{{({{C}_{n{\,^{14}}{\rm{C}}}})}^2}}}$  ratio for the resonance state ${}^{15}{\rm F}(1d_{5/2}^{+})$ and the mirror bound state ${}^{15}{\rm C}(1d_{5/2}^{+})$;  
the solid red line is the  $\frac{{{\Gamma _{p{\,^{14}}{\rm{O}}}}}}{{{{({{C}_{n{\,^{14}}{\rm{C}}}})}^2}}}$  ratio as a function of $R_{ch}$ calculated using Eq. (\ref{GammaANCmirrorratio2});   the blue dashed line  is the  $\frac{{{\Gamma _{p{\,^{14}}{\rm{O}}}}}}{{{{({{C}_{n{\,^{14}}{\rm{C}}}})}^2}}}$ ratio  calculated  as a function of $R_{ch}$ using Eq. (\ref{Ratioapprox2}).}
\label{fig_GANC15F15Cd521}
\end{figure}

 In Fig. \ref{fig_GANC15F15Cd521}  is shown  the ${\Gamma _{{p\,^{14}}{\rm{O}}}}/C_{{n\,^{14}}{\rm{C}}}^2$ ratio calculated using the Wronskian method and  the simplified Eq. (\ref{Ratioapprox2}).  The  Wronskian ratio at $4.0$ fm is  $0.32$ while Eq. (\ref{Ratioapprox2})  gives $0.31$. Both values are very close to the experimental ratio.

\subsection{Comparison of resonance width for $\mathbf{{}^{18}{\rm Ne}(1^{-}) \to {}^{14}{\rm O}(0.0\,{\rm MeV}) + \alpha}$ and mirror ANC for virtual decay $\mathbf{{}^{18}{\rm O}(1^{-}) \to {}^{14}{\rm C}(0.0\,{\rm MeV}) + \alpha}$}

In this section I determine the ratio ${\Gamma _{{\alpha\,^{14}}{\rm{O}}}}/C_{{\alpha\,^{14}}{\rm{C}}}^2$  for the mirror states  ${}^{18}{\rm Ne}(1^{-})$ and ${}^{18}{\rm O}(1^{-})$.
The resonance energy is  $E_{\alpha{}^{14}{\rm O}(0)}=1.038$ MeV  .  
The binding energy  of the bound state  ${}^{18}{\rm O}(1^{-})$  is $\varepsilon_{\alpha{}^{14}{\rm C}}  = 0.027$ MeV.  The resonance width and the ANC of the mirror states are unknown.

\begin{figure}[htbp]  
\includegraphics[width=0.5\textwidth]{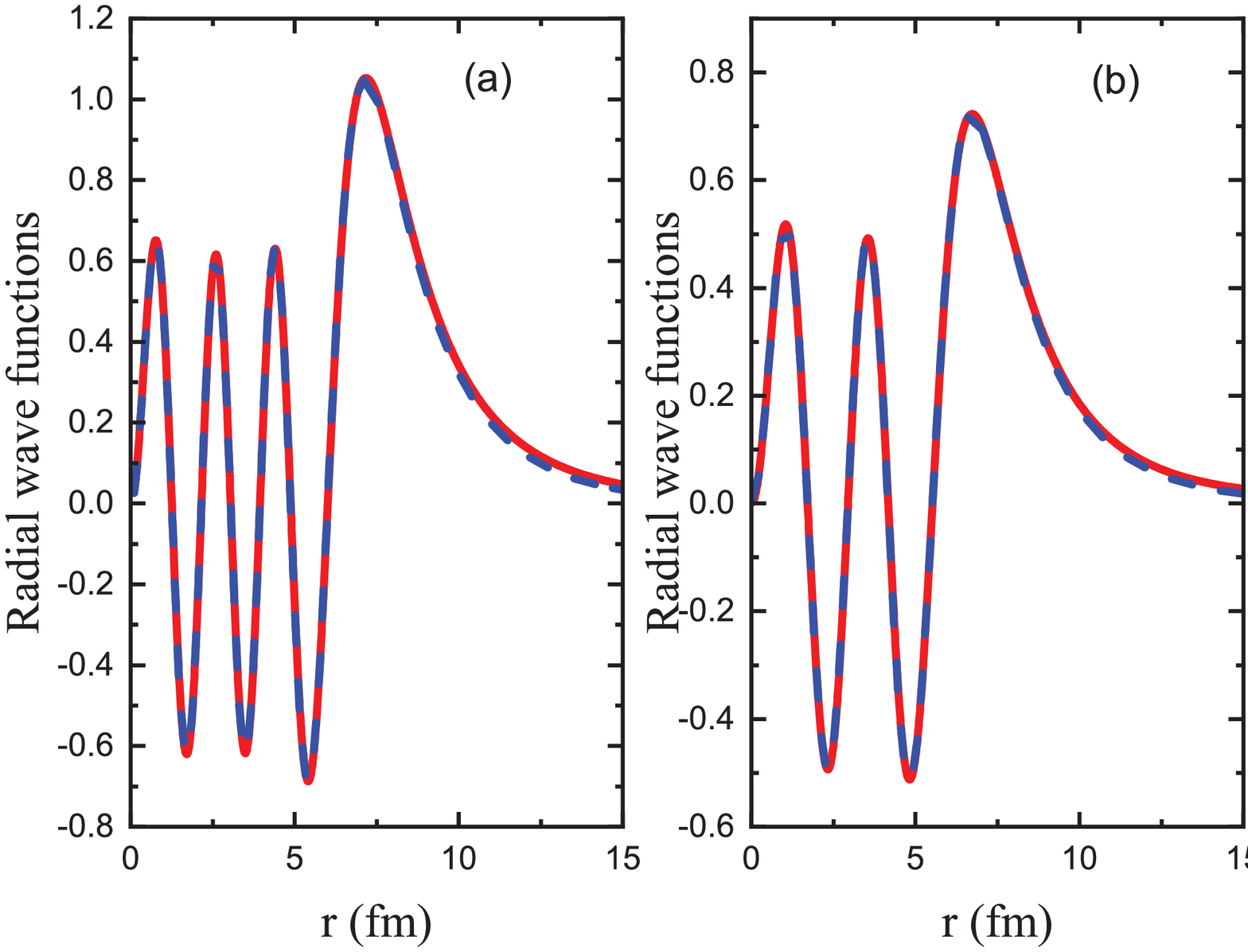}
\caption{ Panel (a): the mirror radial wave functions for $N=6$; the solid red line is the $(\alpha{}^{14}{\rm O})_{1^{-}}$  resonance wave function; the dashed blue line is the radial wave function of  the mirror    $(\alpha{}^{14}{\rm C})_{1^{-}}$  bound-state.  $r$ is the distance between  the $\alpha$-particle  and the c.m. of the nucleus. Panel (b): notations are the same as in panel (a) but for $N=4$.}
\label{fig_radwfs18Ne18O1}
\end{figure}

\begin{figure}[htbp] 
\includegraphics[width=0.4\textwidth]{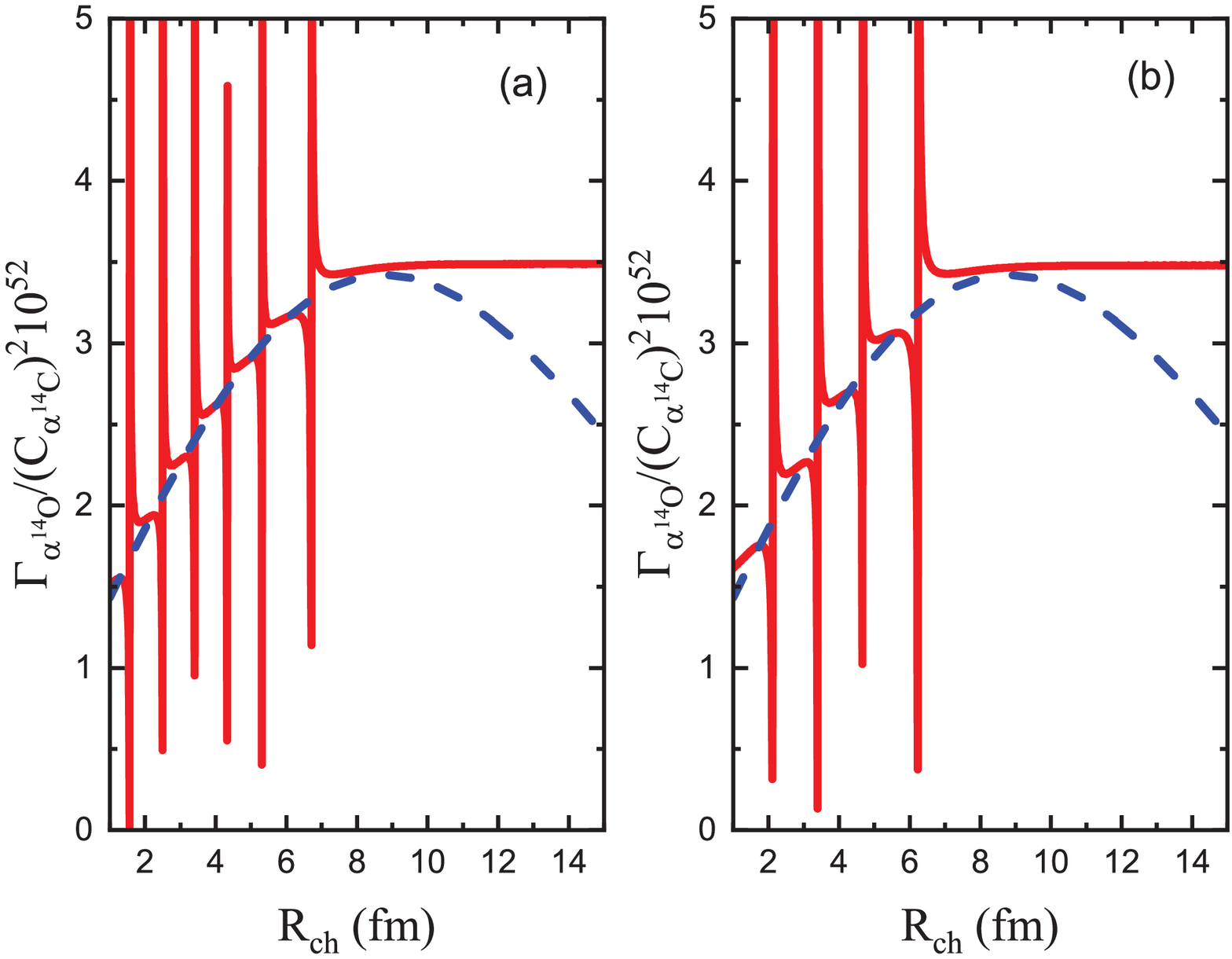}
\caption{Panel (a): the  $\frac{{{\Gamma _{\alpha{\,^{14}}{\rm{O}}}}}}{{{{({{C}_{\alpha{\,^{14}}{\rm{C}}}})}^2}}}$  ratio for the resonance state ${}^{18}{\rm Ne}(1^{-})$ and the mirror bound state ${}^{18}{\rm O}(1^{-})$ for $N=6$;  
the solid red line is the  $\frac{{{\Gamma _{\alpha{\,^{14}}{\rm{O}}}}}}{{{{({{C}_{\alpha{\,^{14}}{\rm{C}}}})}^2}}}$  ratio as a function of $R_{ch}$ calculated using Eq. (\ref{GammaANCmirrorratio2});   the blue dashed line  is the  $\frac{{{\Gamma _{\alpha{\,^{14}}{\rm{O}}}}}}{{{{({{C}_{\alpha{\,^{14}}{\rm{C}}}})}^2}}}$ ratio  calculated  as a function of $R_{ch}$ using Eq. (\ref{Ratioapprox2}). Panel (b): notations are the same as in panel (a) but for $N=4$.}
\label{fig_18Ne18Oratio1}
\end{figure}

The purpose of this section is to show that the ratio  ${\Gamma _{{\alpha\,^{14}}{\rm{O}}}}/C_{{\alpha\,^{14}}{\rm{C}}}^2$ does not depend on the number of the nodes of the mirror wave functions. The potential model search showed that for the given resonance energy and binding energy for $l=1$ the mirror wave functions have 4 or 6 nodes at $r>0$.  
In Figs  \ref{fig_radwfs18Ne18O1} and  \ref{fig_18Ne18Oratio1}   are shown the radial wave functions and the ratio ${\Gamma _{{\alpha\,^{14}}{\rm{O}}}}/C_{{\alpha\,^{14}}{\rm{C}}}^2$  for the number of the nodes $N=4$ and $6$.

One can see that the mirror wave functions practically coincide up to $r=15$ fm. It means that  the simplified Eq. (\ref{Ratioapprox2}) can be used up to $15$ fm.
 The ratio   ${\Gamma _{{\alpha\,^{14}}{\rm{O}}}}/C_{{\alpha\,^{14}}{\rm{C}}}^2$  calculated using Eq. (\ref{GammaANCmirrorratio2}) is the same for $N=4$ and $6$. Because the mirror wave functions practically identical in the external region the ratio 
 ${\Gamma _{{\alpha\,^{14}}{\rm{O}}}}/C_{{\alpha\,^{14}}{\rm{C}}}^2$ calculated using the Wronskian method (Eq. (\ref{GammaANCmirrorratio2}))   has an asymptote. 
 The calculated  for $N=4,\,6$ ratio reaches its asymptotic value at $R_{ch}=7.5$ fm which is  ${\Gamma _{{\alpha\,^{14}}{\rm{O}}}}/C_{{\alpha\,^{14}}{\rm{C}}}^2= 3.48\,\times10^{52}$. 
 The maximum of  ${\Gamma _{{\alpha\,^{14}}{\rm{O}}}}/C_{{\alpha\,^{14}}{\rm{C}}}^2$  calculated using Eq. (\ref{Ratioapprox2})  at $R_{ch}= 9$ fm is $3.42\,\times10^{52}$. This comparison demonstrates again  that in the absence  of the microscopic internal overlap functions both the Wronskian and the simplified method given by Eq.  (\ref{Ratioapprox2}) 
 can be used and give  very close results.

\subsection{Comparison of resonance width for $\mathbf{{}^{17}{\rm F}(s_{1/2}) \to {}^{13}{\rm N}(0.0\,{\rm MeV}) + \alpha}$ and mirror ANC for virtual decay $\mathbf{{}^{17}{\rm O}(s_{1/2}) \to {}^{13}{\rm C}(0.0\,{\rm MeV}) + \alpha}$}

The last case, which I consider,  is the determination of the ratio  $\frac{{{\Gamma _{\alpha{\,^{13}}{\rm{N}}}}}}{{{{({{C}_{\alpha{\,^{13}}{\rm{C}}}})}^2}}}$ of the resonance state ${}^{17}{\rm F}(1/2^{+})$ and the mirror bound state ${}^{17}{\rm O}(1/2^{+})$. The orbital momentum of the mirror states is $l=1$ and the resonance energy is $E_{\alpha\,{}^{13}{\rm N}(0)}= 0.7371$ MeV \cite{Tilley}.  The location of the state ${}^{17}{\rm O}(1/2^{+})$ is questionable. The excitation energy $E_{x}$ of the state ${}^{17}{\rm O}(1/2^{+} )$ is $6356 \pm 8$ keV \cite{Tilley}. 
Taking into account that the $\alpha-{}^{13}{\rm C}$  threshold is located at $6359.2$ keV one finds that this  $1/2^{+}$ level is the located at $E_{ \alpha{}^{13} {\rm C} }=  -3 \pm 8$ keV, that is, it can be a subthreshold bound state or a resonance \cite{Tilley}. 
This location of the level  ${}^{17}{\rm O}(1/2^{+})$ was adopted in the previous analyses of the direct measurements including the latest one in \cite{Heil}. If this level is the subthreshold bound state, then its reduced width is related to the ANC of this level. However, in a recent paper \cite{Faestermann} it has been determined that this level is actually a resonance located at  $E_{\alpha{}^{13}{\rm C}} = 4.7 \pm 3$ keV.
Because the possible subthreshold state and near threshold resonance are located very close to each other the reduced widths corresponding to these two levels are  very close.
Here in the analysis I still assume that  ${}^{17}{\rm O}(1/2^{+})$ is the bound state with the binding energy  of $-3$ keV.  I adopt the ANC of this subthreshold state
$C_{\alpha{}^{13}{\rm C}}^{2}= 4.4\times10^{169}$ fm$^{-1}$ \cite{muk2017}.  

\begin{figure}[htbp]  
\includegraphics[width=0.5\textwidth]{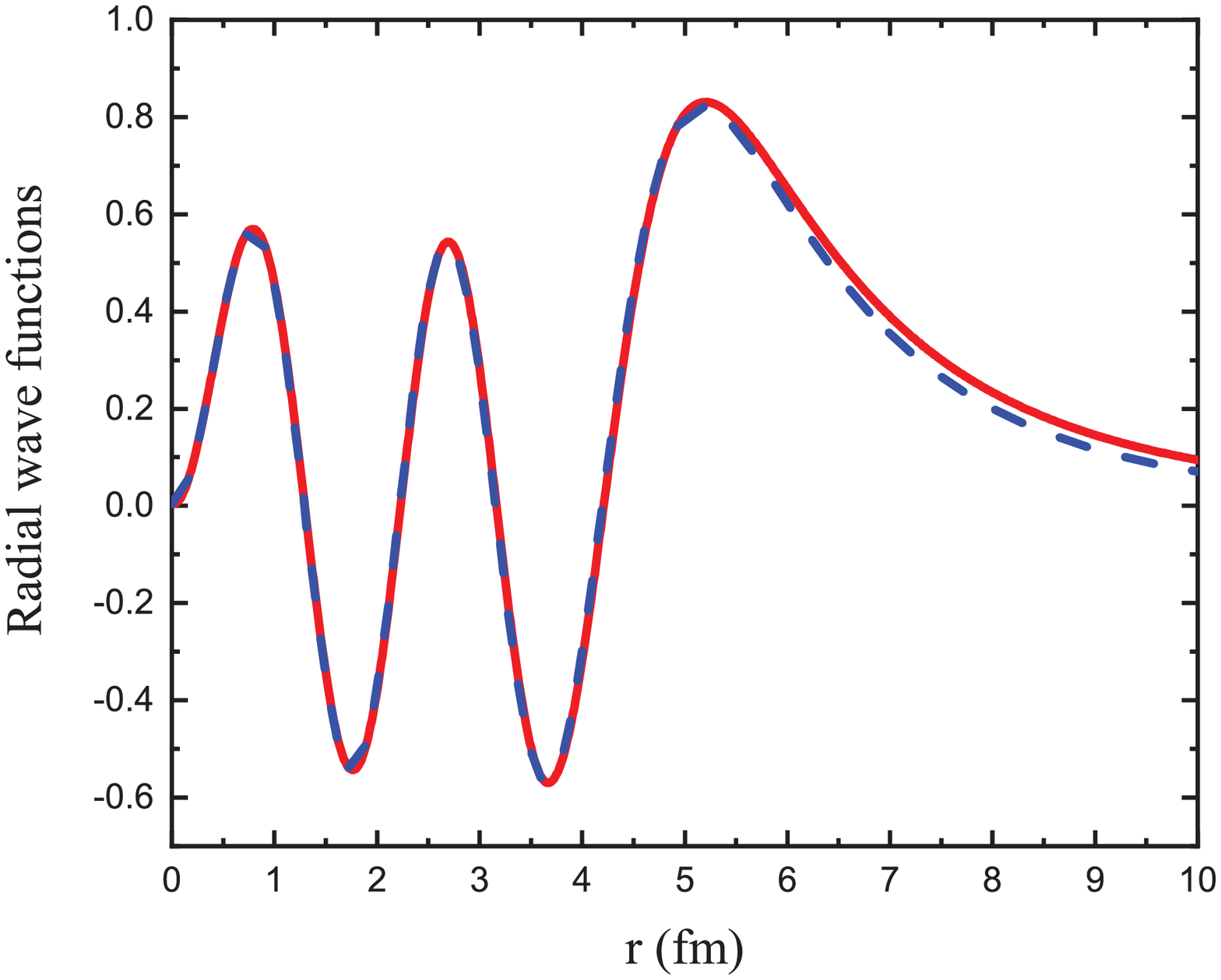}
\caption{ The solid red line is the $(\alpha{}^{13}{\rm N})_{1/2^{+}}$  resonance wave function; the dashed blue line is the radial wave function of  the mirror    $(\alpha{}^{13}{\rm C})_{1/2^{+}}$  bound-state.  $r$ is the distance between  the $\alpha$-particle  and the c.m. of the nucleus.}
\label{fig_radwfs17F17O1}
\end{figure}

The calculated mirror resonance and bound state wave functions are shown in Fig. \ref{fig_radwfs17F17O1}.  Both wave functions practically identical up to $R_{ch} \leq 15$ fm. 

\begin{figure}[htbp] 
\includegraphics[width=0.5\textwidth]{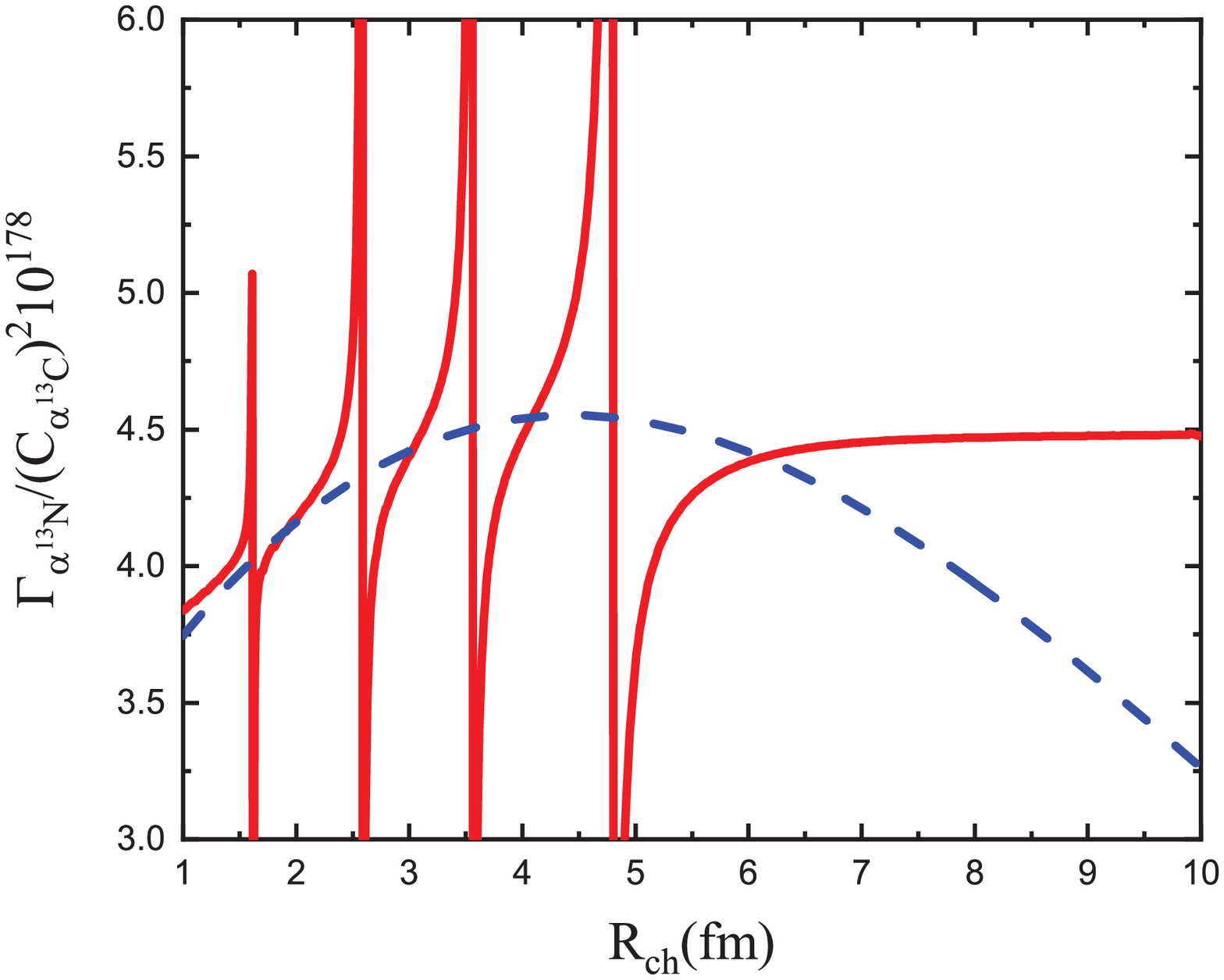}
\caption{The  $\frac{{{\Gamma _{\alpha{\,^{13}}{\rm{N}}}}}}{{{{({{C}_{\alpha{\,^{13}}{\rm{C}}}})}^2}}}$  ratio for the resonance state ${}^{17}{\rm F}(1/2^{+})$ and the mirror bound state ${}^{17}{\rm O}(1/2^{+})$;  
the solid red line is the  $\frac{{{\Gamma _{\alpha{\,^{13}}{\rm{N}}}}}}{{{{({{C}_{\alpha{\,^{13}}{\rm{C}}}})}^2}}}$  ratio as a function of $R_{ch}$ calculated using Eq. (\ref{GammaANCmirrorratio2});   the blue dashed line  is the  $\frac{{{\Gamma _{\alpha{\,^{13}}{\rm{N}}}}}}{{{{({{C}_{\alpha{\,^{13}}{\rm{C}}}})}^2}}}$ ratio  calculated  as a function of $R_{ch}$ using Eq. (\ref{Ratioapprox2}).}
\label{fig_17F17Oratio1}
\end{figure}

In Fig. \ref{fig_17F17Oratio1}  the $\frac{{{\Gamma _{\alpha{\,^{13}}{\rm{N}}}}}}{{{{({{C}_{\alpha{\,^{13}}{\rm{C}}}})}^2}}}$  ratio is calculated using the Wronskian Eq. (\ref{GammaANCmirrorratio2}) and the simple Eq. (\ref{Ratioapprox2}).  The asymptotic value of the ratio is $\frac{{{\Gamma _{\alpha{\,^{13}}{\rm{N}}}}}}{{{{({{C}_{\alpha{\,^{13}}{\rm{C}}}})}^2}}}= 4.48 \times 10^{-178}$. The value of the $\frac{{{\Gamma _{\alpha{\,^{13}}{\rm{N}}}}}}{{{{({{C}_{\alpha{\,^{13}}{\rm{C}}}})}^2}}}$  at the border of  the internal region $R_{ch} = 5.2$  fm is  very close to its asymptotic value.  Eq. (\ref{Ratioapprox2})  gives $\frac{{{\Gamma _{\alpha{\,^{13}}{\rm{N}}}}}}{{{{({{C}_{\alpha{\,^{13}}{\rm{C}}}})}^2}}} =
  4.55 \times 10^{-178}$.  Taking into account the adopted value of the ANC $C_{\alpha\,{}^{13}{\rm C}}$ and the experimental ratio $\frac{{{\Gamma _{\alpha{\,^{13}}{\rm{N}}}}}}{{{{({{C}_{\alpha{\,^{13}}{\rm{C}}}})}^2}}} = 4.48 \times  10^{-178}$   one obtains  from the Wronskian ratio the resonance width 
  $\Gamma_{\alpha\,{}^{13}N} =  4.48 \times  10^{-178} \times 4.4\times10^{169}\,\times \hbar\,c= 3.9 $ eV.

\section{Appendix}
In this Appendix is shown  that Zeldovich regularization procedure can be used for normalization of the resonance wave function  $u_{k_{p}l_{B}}(r)$  both for exponentially decaying potentials and potentials with the Coulomb tail. The normalization of the resonance wave function depends on its tail. Taking into account Eq. (\ref{radcondres1})  it is enough to consider
the integral
\begin{align}
I(\beta, \nu,z) =  \int\limits_{0}^{\infty}{\rm d}r\,e^{-\beta\,r^{2}}\,e^{z\,r}\,r^{\nu}.
\label{Intappend1}
\end{align}
Here,    $\;z=2\, i\,k_{aA(R)}\,r= 2\, i\,k_{aA(0)}\,r + 2\,{\rm Im}{k_{aA(R)}}\,r$. It assumed that $k_{aA(0)} > {\rm Im}{k_{aA(R)}}$, as it should be for physical resonances. 
Then ${\rm Re}z^{2} <0$.  Also 
\begin{align}
\nu = - 2\,i\,\eta_{aA}^{R}= -2i\, \frac{\gamma}{ k_{aA(0)} - i\,{\rm Im}\,{k_{aA(R) }} } = -2\,i\,\frac{  \gamma\,(k_{aA(0)}} { k_{aA(0)}^{2} +   ({\rm Im}{k_{aA(R)} } )^{2}}
+ 2\, \frac{ \gamma\,{\rm Im}{ k_{aA(R)} }     }    { k_{aA(0)}^{2} +   ({\rm Im}{  k_{aA(R) } })^{2}},
\label{nu1}
\end{align}
$\gamma = Z_{a}\,Z_{A}\,\mu_{aA}/137$. Thus, one can see that  for the repulsive Coulomb potential ${\rm Re} \nu >0$.  

Using Eq.  (3.462.1) from  \cite{GradRyzhik}   one gets 
\begin{align}
I(\beta, \nu,z)= \Gamma(\nu+1)\,(2\,\beta)^{-(\nu+ 1)/2}\,e^{z^{2}/(8\,\beta)} \, D_{-\nu -1}(-z/\sqrt{2\,\beta}),
\label{Ibnz1}
\end{align}
where $D_{\sigma}(x)$ is the parabolic cylinder function.  For ${\rm Re}z^{2} <0$  using Eq.  (9.246.1) from \cite{GradRyzhik} one gets
\begin{align}
I(0,\, \nu,\,z) = \lim_{\beta \to +0} \,I(\beta,\,\nu,\,z)  =  \Gamma(\nu+1)\,(-z)^{-\nu-1}.
\label{Ionuz1}
\end{align}
Thus the regularization procedure used by Zeldovich is applicable and for the physical resonances  $\,k_{aA(0)} > {\rm Im}{k_{aA(R)}}\,$  the integral in Eq. (\ref{Intappend1}) does exist and converges in the $\lim\,{\beta \to  + 0}$ . 

Let me consider now the integral
\begin{align}
I_{R}(\beta, \nu,z) =  \int\limits_{R}^{\infty}{\rm d}r\,e^{-\beta\,r^{2}}\,e^{z\,r}\,r^{\nu}.
\label{IntappendR1}
\end{align}
Integrating it by parts one gets
\begin{align}
\lim_{\beta \to +0}\,I_{R}(\beta, \nu,z) =  -\frac{R^{\nu}}{z}\,e^{z\,R}\,\Big[ 1 - \frac{\nu}{z\,R} + O(\frac{1}{z^{2}\,R^{2}}) \Big].
\label{IntappendR1}
\end{align}

\section{acknowledgments}  
This work was supported by the U.S. DOE Grant No. DE-FG02-93ER40773, NNSA Grant No. DE-NA0003841 and U.S. NSF Award No. PHY-1415656.

\end{document}